%% file: PFT-25-001_temp.tex
\pdfoutput=1
\documentclass[11pt,twoside,a4paper,cmspaper,final,collab]{cms-tdr}
\def\svnVersion{d88da74}\def\svnDate{2026/06/24}\def\cmsCernNoTag{CERN-EP-2025-300}\def\cmsCernDate{\today}\def\cmsMessage{Published in the European Physical Journal C as \href{http://dx.doi.org/10.1140/epjc/s10052-026-15754-5}{\doi{10.1140/epjc/s10052-026-15754-5}.}}
\begin{document}\cmsNoteHeader{PFT-25-001}

\newlength\cmsTabSkip\setlength{\cmsTabSkip}{1ex}
\ifthenelse{\boolean{cms@external}}{\providecommand{\cmsLeft}{upper\xspace}}{\providecommand{\cmsLeft}{left\xspace}}
\ifthenelse{\boolean{cms@external}}{\providecommand{\cmsRight}{lower\xspace}}{\providecommand{\cmsRight}{right\xspace}}
\newcommand{\pthat}{\ensuremath{\hat{p}_\mathrm{T}}\xspace}
\newcommand{\qcd}{QCD\xspace}
\newcommand{\Ztautau}{\ensuremath{\PZ \to \PGt\PGt}\xspace}
\newcommand{\PYTORCH} {{\textsc{PyTorch}}\xspace}
\newcommand{\FlashAttention} {{\textsc{FlashAttention}}\xspace}
\newcommand{\ONNX} {{\textsc{ONNX}}\xspace}
\newcommand{\ONNXRUNTIME} {{\textsc{onnxruntime}}\xspace}
\newcommand{\ptPtcl}{\ensuremath{p_{\text{T,ptcl}}}\xspace}
\newcommand{\ptRaw}{\ensuremath{p_{\text{T,raw}}}\xspace}
\newcommand{\ptCorr}{\ensuremath{p_{\text{T,corr}}}\xspace}
\hyphenation{cal-or-i-me-ter}

\cmsNoteHeader{PFT-25-001}

\title{Full event interpretation with machine-learning-based particle-flow reconstruction in the CMS detector}

\date{\today}

\abstract{The particle-flow (PF) algorithm constructs a global description of each particle collision by producing a comprehensive list of final-state particles, and is central to event reconstruction in the CMS experiment at the CERN LHC.
    The existing PF implementation relies on physics-motivated heuristics and assumptions that can be replaced by machine-learning (ML) models trained directly on simulated data and naturally suited to modern graphics processing units (GPUs). A state-of-the-art ML-based PF (MLPF) reconstruction algorithm, implemented within the CMS software framework, is presented. The MLPF algorithm performs a learnable full-event reconstruction on GPUs, generalizes across detector conditions and collision energies, and replaces multiple modular reconstruction steps with a single unified model. Physics performance comparable to standard PF reconstruction is achieved in both simulation and data, with improved jet energy resolution and inference time. In simulated top quark-antiquark events under LHC Run-3 (2023--2024) conditions, the jet energy resolution improves by 10--20\% for jets with transverse momentum between 30--100\GeV.Inference time is evaluated using simulated multijet events, with a median of 20\unit{ms} per event on an Nvidia L4 GPU, compared to approximately 110\unit{ms} for the standard CMS PF reconstruction.
}

\hypersetup{
    pdfauthor={CMS Collaboration},
    pdftitle={Full event interpretation with machine-learning-based particle-flow reconstruction in the CMS detector},
    pdfsubject={CMS},
    pdfkeywords={CMS, machine learning, particle-flow, event reconstruction, inference}
}

\maketitle

\section{Introduction}
\label{sec:intro}

The particle-flow (PF) algorithm aims to reconstruct a global, particle-level view of each event by combining information from multiple subdetectors in general-purpose particle detectors.
By reconciling the signals associated with each particle across the tracker, calorimeters, and muon systems, PF removes ambiguities, establishes correlations between subdetector signatures, and produces an optimized, self-consistent list of reconstructed particles, classified as photons, electrons, muons, charged hadrons, or neutral hadrons.
This approach significantly improves the event description and the resolution of high-level reconstructed quantities---commonly referred to as physics objects---such as jets, missing transverse momentum (\ptvecmiss), photons, and leptons, compared to methods relying on individual subdetector systems.
The PF algorithm has been central to CMS analyses~\cite{CMS:2017yfk} since LHC Run~1 (2009--2013).
Similar strategies have been adopted at other experiments, including ALEPH~\cite{Buskulic:1994wz}, ATLAS~\cite{Aaboud:2017aca}, CDF~\cite{Bocci:2001zx,Connolly:2003gb,Abulencia:2007iy}, CELLO~\cite{Behrend:1982gk}, CLIC~\cite{Marshall:2012ry}, D0~\cite{Abazov:2008ff}, DELPHI~\cite{Abreu:1995uz}, H1~\cite{H1:2020zpd}, ILC~\cite{Thomson2009}, and ZEUS~\cite{Breitweg:1997aa,Breitweg:1998gc}.
The PF concept is also central to the design of the proposed lepton collider detectors~\cite{FCC-ee,Behnke2013,CEPCStudyGroup2018}, where detector parameters are still being optimized.
In that context, approaches to PF based on machine learning (ML) can significantly accelerate detector design studies by enabling efficient retraining and rapid iteration, as demonstrated in Ref.~\cite{Mokhtar:2025mlpf}.

Current production-level event reconstruction algorithms typically employ iterative, rule-based methods, such as proximity-based linking, where tracks and calorimeter clusters are associated based on geometric closeness in detector coordinates.
In contrast, ML approaches to particle identification are particularly well-suited to irregular, high-granularity detector geometries.
While early ML-based event reconstruction algorithms relied on computer vision techniques~\cite{DiBello:2020bas}, more recent work has shifted toward graph neural networks (GNNs)~\cite{Kieseler:2020wcq,Shlomi:2020gdn,Pata:2021oez,CMS:2022txh,Pata:2022wam,Mokhtar:2023fzl}, hypergraphs using clustering~\cite{DiBello:2022iwf}, and transformer architectures~\cite{Pata:2023rhh,Wahlen:2024rxt,Kakati:2024dun,Kobylianskii:2025nrj}, demonstrating scalability and accuracy in complex event reconstruction.
In addition, advances in modern computer hardware, particularly graphics processing units (GPUs), together with ML architectures, such as GNNs and transformers, enable efficient scaling to high-dimensional, variable-size input and output spaces~\cite{CMS:2024twn}, a capability that is critical for event reconstruction.

Current active experiments, such as CMS, offer an excellent platform for developing ML-based event reconstruction techniques in parallel with the existing PF algorithms.
These efforts can be validated directly in data, with continuous feedback from CMS analysis working groups.
If successful, they have the potential to drive the design and deployment of algorithms for the high-luminosity upgrade of the CERN LHC (HL-LHC)~\cite{Aberle:2749422} and future colliders, ultimately enhancing the physics reach of the experiments and paving the way toward a detector optimization pipeline with differentiable programming~\cite{Dorigo:2023E2EOpt,Aehle:2025E2EProgress}.
Furthermore, ML-based reconstruction algorithms offer the possibility of incorporating analysis-specific information or selection biases directly into the reconstruction procedure, either through fine-tuning or dedicated retraining.
For example, ML techniques were employed in the CMS measurement of the \PW boson mass to improve the resolution of the reconstructed \ptvecmiss, resulting in an enhanced separation between signal and background~\cite{CMS:2024lrd}.
Such flexibility is difficult to achieve with traditional PF algorithms, but becomes feasible in an ML-based framework.

Over the past decade, ML has been widely adopted at the LHC, enabled by advances in GPU hardware and neural network architectures.
Jet flavor tagging has improved significantly, with models such as \textsc{ParticleNet}~\cite{Qu:2019gqs} and \textsc{ParT}~\cite{Qu:2022mxj} achieving 2--5 times better jet identification compared to previous baselines~\cite{Moreno:2019bmu,Moreno:2019neq,Mikuni:2020wpr}.
Machine-learning methods have also been applied to tracking~\cite{Farrell:2018cjr,Ju:2020xty,Amrouche:2019wmx,Amrouche:2019yxv,Choma:2020cry,Dezoort:2021kfk,VanStroud:2024fau,EPJCTDL:2017trk,Ju:2021geometric,Miao:2024oqy}, secondary vertex reconstruction~\cite{Shlomi:2020ufi,VanStroud:2023ggs}, jet clustering~\cite{Ju:2020tbo,Li:2020grn,Guo:2020vvt}, jet assignment~\cite{Fenton:2020woz,Shmakov:2021qdz,Fenton:2023ikr,Li:2024qfq}, calorimeter reconstruction~\cite{Qasim:2019otl}, and pileup mitigation~\cite{Komiske:2017ubm,ArjonaMartinez:2018eah,Li:2022omf}, where pileup refers to the number of additional proton--proton ($\Pp\Pp$) interactions occurring in the same bunch crossing as the hard scattering process.

The goal of this paper is to explore and establish ML-based approaches to global event reconstruction that improve physics performance and computational scalability on GPUs, while laying the groundwork for event reconstruction at the HL-LHC.
At the HL-LHC, pileup is expected to reach up to 200 on average.
To operate under these conditions, the CMS detector is undergoing a comprehensive Phase-2 upgrade program, including major improvements to the tracking, trigger, and calorimeter systems~\cite{CERN-LHCC-2017-009,CERN-LHCC-2019-003,CERN-LHCC-2020-004,CERN-LHCC-2021-007,CERN-LHCC-2021-008,CERN-LHCC-2017-011,CMS:2017dhl,CERN-LHCC-2017-012}.
As reconstruction complexity continues to grow with increasing pileup, there is strong motivation to incorporate ML-based pattern recognition techniques in the reconstruction chain.
While this paper is limited to Run~3, this approach may be extended to HL-LHC conditions, where given the increased complexity of the events, we expect further benefits from ML-based solutions.

    {\tolerance=1200
        In this paper, we present the implementation of an ML-based PF (MLPF) reconstruction algorithm within the CMS experiment.
        This is the first MLPF reconstruction model that has been applied to data collected by a hadron collider experiment.
        The model demonstrates improved performance over the standard PF while maintaining practical inference throughput on GPUs in production software.
        We begin by introducing the CMS detector (Section~\ref{sec:cms}) and the current reconstruction algorithms (Section~\ref{sec:baseline_reco}), followed by a description of the data sets used in this work (Section~\ref{sec:datasets}).
        We then define the reconstruction-level target (Section~\ref{sec:target}), designed to closely align with the generator, particle-level Monte Carlo (MC) information.
        This well-defined and physically meaningful target enables the application of supervised learning for training the MLPF algorithm.
        We then describe the architecture of the algorithm (Section~\ref{sec:mlpfalgo}), which is based on transformer networks trained end-to-end on MC simulation samples, along with the associated training procedure and multi-task loss function.
        The physics performance of the algorithm is evaluated in Section~\ref{sec:perf_sim}.
        In addition, the MLPF algorithm has been successfully applied in data collected by the CMS detector during Run~3 (2023--2024) (Section~\ref{sec:perf_data}), marking the first data-validated application of an ML-based reconstruction pipeline at any LHC experiment.
        In Section~\ref{sec:cmssw}, we assess the computational performance and compare with the standard PF algorithm.
        Finally, Section~\ref{sec:sum} summarizes the paper.
        \par}

\section{The CMS detector}
\label{sec:cms}

The CMS apparatus~\cite{CMS:2008xjf,CMS:2023gfb} is a multipurpose, nearly hermetic detector, designed to trigger on~\cite{CMS:2020cmk,CMS:2016ngn,CMS:2024aqx} and identify electrons, muons, photons, charged and neutral hadrons~\cite{CMS:2020uim,CMS:2018rym,CMS:2014pgm}.
Its central feature is a superconducting solenoid of 6\unit{m} internal diameter, providing a magnetic field of 3.8\unit{T}.
Within the solenoid volume are a silicon pixel and strip tracker, a lead tungstate crystal electromagnetic calorimeter (ECAL), and a brass and scintillator hadron calorimeter (HCAL), each composed of a barrel and two endcap sections.
Forward calorimeters extend the pseudorapidity ($\eta$) coverage provided by the barrel and endcap detectors.
Muons are reconstructed using gas-ionization detectors embedded in the steel flux-return yoke outside the solenoid.
More detailed descriptions of the CMS detector, together with a definition of the coordinate system used and the relevant kinematic variables, can be found in Refs.~\cite{CMS:2008xjf,CMS:2023gfb}.

Events of interest are selected using a two-tiered trigger system.
The first level, composed of custom hardware processors, uses information from the calorimeters and muon detectors to select events at a rate of around 100\unit{kHz} within a fixed latency of 4\mus~\cite{CMS:2020cmk}.
The second level, known as the high-level trigger (HLT), consists of a farm of processors running a version of the full event reconstruction software optimized for fast processing, and reduces the event rate to a few kHz before data storage~\cite{CMS:2016ngn,CMS:2024aqx}.

\section{Reconstruction algorithms}
\label{sec:baseline_reco}

Reconstruction algorithms in CMS are generally categorized as either local or global.
Local reconstruction algorithms operate within individual detector subsystems, such as track reconstruction in the tracker or clustering in the calorimeters.
Global reconstruction algorithms, such as PF, combine information across multiple subsystems to reconstruct higher-level physics objects.

{\tolerance=1200
Although the reconstruction algorithms used in CMS span different detector subsystems, they share a common computational structure: clustering or pattern recognition based on irregular, variable-size collections of detector hits.
Prior to the use of ML algorithms, such clustering tasks were typically solved with hand-engineered methods whose naive computational complexity scales as $\mathcal{O}(N^3)$, where $N$ denotes the number of detector hits (or clusters) under consideration, when searching for compatible hit triplets, or clusters, unless strong inductive biases are introduced to restrict the search space (often to $\mathcal{O}(N\log{N})$).
These complexities arise naturally in particle physics reconstruction, where the data are unstructured and the combinatorics grow rapidly with detector occupancy.
Historically, this has often limited the applicability of standard computer science clustering algorithms to LHC tasks, leading experiments to repeatedly design custom algorithms tuned to detector-specific features, an approach that is not ideal for long-term scalability.
Modern ML architectures---most notably GNNs and transformers---provide new, expressive inductive biases that operate natively on sets or graphs, enabling more efficient representation learning and reducing the need for handcrafted heuristics.
This evolution opens the door to reconstruction approaches that scale more favorably with event complexity while maintaining physics fidelity.
With this context in mind, we now briefly describe the reconstruction algorithms currently used in CMS.
\par
}

\subsection{Track reconstruction}
{\tolerance=1200
    Reconstructing charged-particle tracks from detector hits is a complex combinatorial problem.
    Current state-of-the-art methods in CMS rely on the combinatorial track finder, which has evolved from combinatorial Kalman filters.
    Track reconstruction proceeds iteratively, beginning with seed generation, in which initial track candidates are identified using a small number of hits (typically 3)~\cite{BILLOIR1990219}, followed by track finding, where hits are grouped into candidate tracks.
    The track parameters are then determined using a fit to the associated hits, and finally, quality criteria are applied during a filtering stage to retain high-quality tracks.
    \par}

Each iteration targets progressively more challenging tracks, removing hits associated with previously reconstructed tracks to reduce combinatorial complexity.
The specific parameters for each stage---finding, fitting and filtering---are tunable and optimized for specific data-taking periods.
A detailed description of the CMS track reconstruction algorithm can be found in Ref.~\cite{CMS:2014pgm}.
The track trajectory, constituent hits, and associated quality criteria derived from the tracking system are used as input to both the PF and MLPF algorithms.

\subsection{Calorimeter clustering}

The objective of calorimeter clustering in PF is to group energy deposits from calorimeter hits originating from the same particle.
Clustering is performed independently in the HCAL, ECAL, and the hadron forward (HF) calorimeter, whose long and short fiber readouts are combined to form the HF hadronic and electromagnetic sections.
The current PF clustering algorithm, which uses seeding followed by an expectation-maximization Gaussian mixture model, has already been ported to GPUs~\cite{CMS-DP-2024-026}.
The existing PF clusters are used as inputs to MLPF, making the MLPF approach fully complementary to ongoing GPU-based developments in PF clustering.
For both PF and MLPF reconstruction, the cluster energy and position are used as inputs.
Additionally, in Section~\ref{sec:inp_features} we compute detailed per-cluster features based on the energy-weighted per-layer distribution of the hits, demonstrating that the MLPF approach can efficiently incorporate low-level information in the PF algorithm.

\subsection{Particle-flow reconstruction}

The global event reconstruction, also called PF reconstruction, aims to reconstruct and identify each individual particle in an event, with an optimized combination of all subdetector information.
In this process, the identification of the particle type (photon, electron, muon, charged hadron, or neutral hadron) plays an important role in the determination of the particle direction and energy.
Photons (\eg, coming from \PGpz decays or from electron bremsstrahlung) are identified as ECAL energy clusters not linked to the extrapolation of any charged-particle trajectory to the ECAL.
Electrons (\eg, produced promptly in the hard interaction or arising from secondary processes such as photon conversions in the tracker material or semileptonic \PQb hadron decays) are identified as a primary charged-particle track and potentially many ECAL energy clusters corresponding to this track extrapolation to the ECAL and to possible bremsstrahlung photons emitted along the way through the tracker material.

Muons (\eg, from \PQb hadron semileptonic decays) are identified as tracks in the central tracker consistent with either a track or several hits in the muon system, and associated with calorimeter deposits compatible with the muon hypothesis.
Charged hadrons are identified as charged-particle tracks neither identified as electrons, nor as muons.
Finally, neutral hadrons are identified as HCAL energy clusters not linked to any charged-hadron trajectory, or as a combined ECAL and HCAL energy excess with respect to the expected charged-hadron energy deposit.

The energy of photons is obtained from the ECAL measurement.
The energy of electrons is determined from a combination of the track momentum at the main interaction vertex, the corresponding ECAL cluster energy, and the energy sum of all bremsstrahlung photons attached to the track.
At low transverse momentum (\pt), the energy of muons is obtained from the corresponding track momentum, while at high \pt, the PF algorithm evaluates several candidate fits and selects the one with the best quality among the tracker-only fit, the fit using the tracker combined with the first muon station, the full global-muon fit, and a global-muon fit excluding muon chambers affected by high occupancy.
The energy of charged hadrons is determined from a combination of the track momentum and the corresponding ECAL and HCAL energies, corrected for the response function of the calorimeters to hadronic showers.
Finally, the energy of neutral hadrons is obtained from the corresponding corrected ECAL and HCAL energies.

\subsubsection{Jet and \texorpdfstring{\ptmiss}{pTmiss} reconstruction}

{\tolerance=1200
    Once the PF algorithm has reconstructed the final-state particle candidates, higher-level reconstruction steps, including jet reconstruction, are performed.
    For each event, hadronic jets are clustered from these reconstructed particles using the infrared- and collinear-safe anti-\kt algorithm~\cite{Cacciari:2008gp, Cacciari:2011ma} with a distance parameter of 0.4 (or 0.8).
    Jet momentum is determined as the vectorial sum of all particle momenta in the jet, and is found from simulation to be, on average, within 5--10\% of the true momentum for $\pt>30\GeV$~\cite{CMS:2016lmd}.
    For the purposes of determining jet momentum response, the true particle-level \pt of the jet, excluding neutrinos, is denoted as \ptPtcl, and includes only particles produced from the hard scatter.
    The \pt of jets reconstructed from the output particles of PF or MLPF, before any jet-level corrections, is denoted as \ptRaw.
    The raw jet response is defined as the ratio $\ptRaw / \ptPtcl$.
    Jet energy corrections are derived from simulation to bring the measured response of jets to that of particle-level jets on average.
    The jet $\pt$ is corrected as a function of $\ptRaw$ and $\eta$ by extracting the Gaussian mean or arithmetic median and standard deviation of the raw response distribution in bins of $\ptPtcl$ and $\eta$.
    The corrected \pt of the jet is denoted as $\ptCorr$, and is used to define the corrected jet response $\ptCorr / \ptPtcl$.
    In general, in situ measurements of the momentum balance in dijet, $\gamma$+jet, $\PZ$+jet, and multijet events can be used to account for any residual differences in the jet energy scale between data and simulation~\cite{CMS:2016lmd}.
    The jet energy resolution amounts typically to 15--20\% at 30\GeV, 10\% at 100\GeV, and 5\% at 1\TeV~\cite{CMS:2016lmd}.
    Additional selection criteria are applied to each jet to remove jets potentially dominated by anomalous contributions from various subdetector components or reconstruction failures.
    In this result, for the sake of comparing the PF and MLPF algorithms, the simulated response of jets is calibrated~\cite{CMS:2016lmd}, extracting the simulation-based corrections from events composed uniquely of jets produced through the strong interaction, known as quantum chromodynamics (\qcd) multijet events, in bins of jet $\eta$ and \ptPtcl, and applying them to all samples uniformly using a two-dimensional linear interpolation.
    Pileup can contribute additional tracks and calorimetric energy depositions to the jet momentum.
    The pileup-per-particle identification algorithm (PUPPI)~\cite{Sirunyan:2020foa,Bertolini:2014bba} is used to mitigate the effect of pileup at the reconstructed-particle level, making use of local shape information, event pileup properties, and tracking information.
    A local shape variable is defined, which distinguishes between collinear and soft diffuse distributions of other particles surrounding the particle under consideration.
    The former is attributed to particles originating from the hard scatter and the latter to particles originating from pileup interactions.
    Charged particles identified to be originating from pileup vertices are discarded from the jet momentum measurement.
    For each neutral particle, a local shape variable is computed using the surrounding charged particles compatible with the primary vertex (PV) within the tracker acceptance ($\abs{\eta} < 2.5$), where the PV is taken to be the vertex corresponding to the hardest scattering in the event, evaluated using tracking information alone, as described in Section 9.4.1 of Ref.~\cite{CMS-TDR-15-02}.
    Outside the tracker coverage, both charged and neutral particles are used.
    The momenta of the neutral particles are then rescaled according to their probability to originate from the PV deduced from the local shape variable, superseding the need for jet-based pileup corrections~\cite{Sirunyan:2020foa}.
    \par}

The vector \ptvecmiss is defined as the corrected negative vector \pt sum of all the PF candidates in an event, and its magnitude is denoted as \ptmiss~\cite{CMS:2019ctu}.
The \ptvecmiss is computed from the PF candidates weighted by their probability to originate from the PV~\cite{CMS:2019ctu}, and is modified to account for corrections to the energy scale of the reconstructed jets in the event.
The PUPPI algorithm is applied to reduce the pileup dependence of the \ptvecmiss observable.
Anomalous high-\ptmiss events can be due to a variety of reconstruction failures, detector malfunctions or noncollision backgrounds.
Such events are rejected by event filters that are designed to identify more than 85--90\% of the spurious high-\ptmiss events with a mistagging rate less than 0.1\%~\cite{CMS:2019ctu}.

\section{Data sets and simulated samples}
\label{sec:datasets}

Simulated $\Pp\Pp$ collision events for training and validation are generated under Run-3 (2023) detector and accelerator conditions at a center-of-mass energy of $\sqrt{s}=14\TeV$.
Testing events are generated under Run-3 (2024) detector and accelerator conditions at a center-of-mass energy of $\sqrt{s}=13.6\TeV$, demonstrating the out-of-distribution generalizability of MLPF across detector conditions and collision energies.
Parton showering, fragmentation, and hadronization are modeled with \PYTHIA (v8.311)~\cite{Bierlich:2022pfr}, using the CP5 underlying event tune~\cite{Sirunyan:2019dfx}.
The resulting stable particles are propagated through a detailed simulation of the CMS detector response using \GEANTfour (v11.2.2)~\cite{GEANT42003,GEANT42006,GEANT42016}.

The MC simulation samples used for optimizing the MLPF model are summarized in Table~\ref{tab:samples}.
The training and validation samples are divided using an 80\%/20\% split, respectively.
All benchmark performance metrics in simulation, shown in Section~\ref{sec:perf_sim} and onwards, are evaluated using the testing events.
For samples with pileup, the number of additional interactions per bunch crossing is drawn from a uniform distribution between 55--75.
In particular, the \qcd multijet training sample is generated with a \pthat (\pt of the initial-state partons in the rest frame) restricted to the range [15, 3000]\GeV in \PYTHIA, which is a typical configuration for PF validation.
The top quark-antiquark (\ttbar) sample is generated inclusively across all top quark decay modes.

For commissioning on data, we use a small subset of CMS Run-3 data collected in 2024 at $\sqrt{s} = 13.6\TeV$ and passing a dijet HLT selection.
The trigger requires the presence of a leading (subleading) PF jet with $\pt > 125$ (45)\GeV, and the invariant mass of the dijet system must exceed 1.2\TeV.
We use events where the detector was fully operational, and remove events with known sources of detector noise, such as beam halo events or events with spurious \ptmiss~\cite{CMS:2019ctu}.
Since we are interested in comparing the same data events with two different reconstruction algorithms, we use the same HLT software for both PF and MLPF events, and do not apply additional corrections for trigger efficiency.
The integrated luminosity of the data sample used for commissioning studies is 375\pbinv.

\begin{table*}[ht]
    \centering
    \caption{
        Monte Carlo simulation samples used for optimizing, validating, and testing the model.
        The training and validation samples are divided using an 80\%/20\% split, respectively.
        The symbols \NA in the 3rd column represent samples generated without pileup.
    }
    \begin{tabular}{lcccc}
        Process                          & $\sqrt{s}$ (\TeVns) & Pileup & Events & Purpose              \\
        \hline
        \ttbar                           & 14                  & 55--75 & 500\,k & Training, Validation \\
        \qcd $\pthat \in [15, 3000]\GeV$ & 14                  & 55--75 & 500\,k & Training, Validation \\
        \Ztautau all-hadronic            & 14                  & 55--75 & 500\,k & Training, Validation \\[\cmsTabSkip]
        \ttbar                           & 14                  & \NA    & 5\,M   & Training, Validation \\
        \qcd $\pthat \in [15, 3000]\GeV$ & 14                  & \NA    & 5\,M   & Training, Validation \\
        \Ztautau all-hadronic            & 14                  & \NA    & 5\,M   & Training, Validation \\[\cmsTabSkip]
        \ttbar                           & 13.6                & 55--75 & 100\,k & Testing              \\
        \qcd $\pthat \in [15, 3000]\GeV$ & 13.6                & 55--75 & 500\,k & Testing              \\[\cmsTabSkip]
        \ttbar                           & 13.6                & \NA    & 200\,k & Testing              \\
        \qcd $\pthat \in [15, 3000]\GeV$ & 13.6                & \NA    & 200\,k & Testing              \\
    \end{tabular}
    \label{tab:samples}
\end{table*}

\section{Machine learning for PF reconstruction}

Training a supervised ML model for PF reconstruction requires a carefully defined particle-level target.
This target acts as a proxy for the desired reconstruction output and ultimately sets the upper bound for the performance of the ML model.
Defining a physics-motivated, well-defined target is essential, yet challenging due to the complexity of particle decays and interactions with the detector material.
In $\Pp\Pp$ collisions at the CMS experiment, particles may originate from the hard interaction (primaries), be decay products of other primaries, fall outside the detector acceptance, or arise from pileup.

In the next section, we describe the particle-level training target definition used to train the MLPF algorithm for PF reconstruction in CMS.

\subsection{Particle-level training target}
\label{sec:target}

The MLPF training \emph{target} is constructed to closely approximate the full set of stable particles in the detector from both the hard interaction and pileup.
The hard scattering particles produced by the event generator \PYTHIA are referred to as the \textit{truth}, and is used to measure the quality of reconstruction after pileup subtraction is applied on the reconstructed particles.
The truth is used to construct physics-level observables, such as jets and \ptmiss, and serves as the benchmark against which reconstruction performance is evaluated.
The training target, on the other hand, consists of all particles that are, in principle, reconstructable---\ie, those that interact with the detector and leave measurable signals.
For instance, a generator-level particle that falls outside the detector acceptance is included in the truth but excluded from the target, whereas particles arising from secondary interactions with detector material may appear in the target but not the truth.
In samples with pileup overlay, the truth includes only particles from the primary hard-scattering process, excluding contributions from additional $\Pp\Pp$ interactions.

The goal is to construct a training target for reconstruction that is both well-defined and physically meaningful: each target particle must be reconstructable, without ambiguities from simulated showers or double counting.
In addition, the target should be aligned with the truth---closely reflecting the particle-level information, with minimal energy differences arising solely from detector acceptance.
This ensures that the task solved by the reconstruction model is aligned with the underlying physics and is learnable from the available inputs.
In particular, the transformation from tracks and clusters to the target particles should be simple enough to be approximated effectively by an ML model trained on limited-size simulated samples.

We define the target as the subset of particles that leave hits in the detector either directly or through their decay products.
Particles that fail to register any detectable signal---due to low energy, short lifetimes, or falling outside detector acceptance---are excluded from the target.
This naturally accounts for inefficiencies introduced by finite detector granularity, energy thresholds, and upstream reconstruction algorithms, which this work does not aim to correct.
In practice, this mainly impacts low-\pt photon radiation, which often falls outside detector acceptance or below the energy threshold.
Furthermore, charged hadrons that fail to produce a reconstructed track are treated as neutral hadrons in the training targets.
This treatment ensures that the MLPF algorithm learns realistic particle assignments in cases where tracking inefficiencies occur.

We use information from the \GEANTfour simulation to define a pileup-free target, which includes only the target particles associated with the hard interaction.
Although this is not used for training, it allows us to validate the target definition by comparing the target with the truth when the pileup component has been subtracted.
This is done via a pileup label that is typically binary, with a value of zero (one) for particles originating from the hard interaction (pileup).
In cases where a target particle was split or merged, we compute an energy-weighted pileup fraction per target particle.

\begin{figure*}[htbp]
    \centering
    \includegraphics[width=0.38\linewidth]{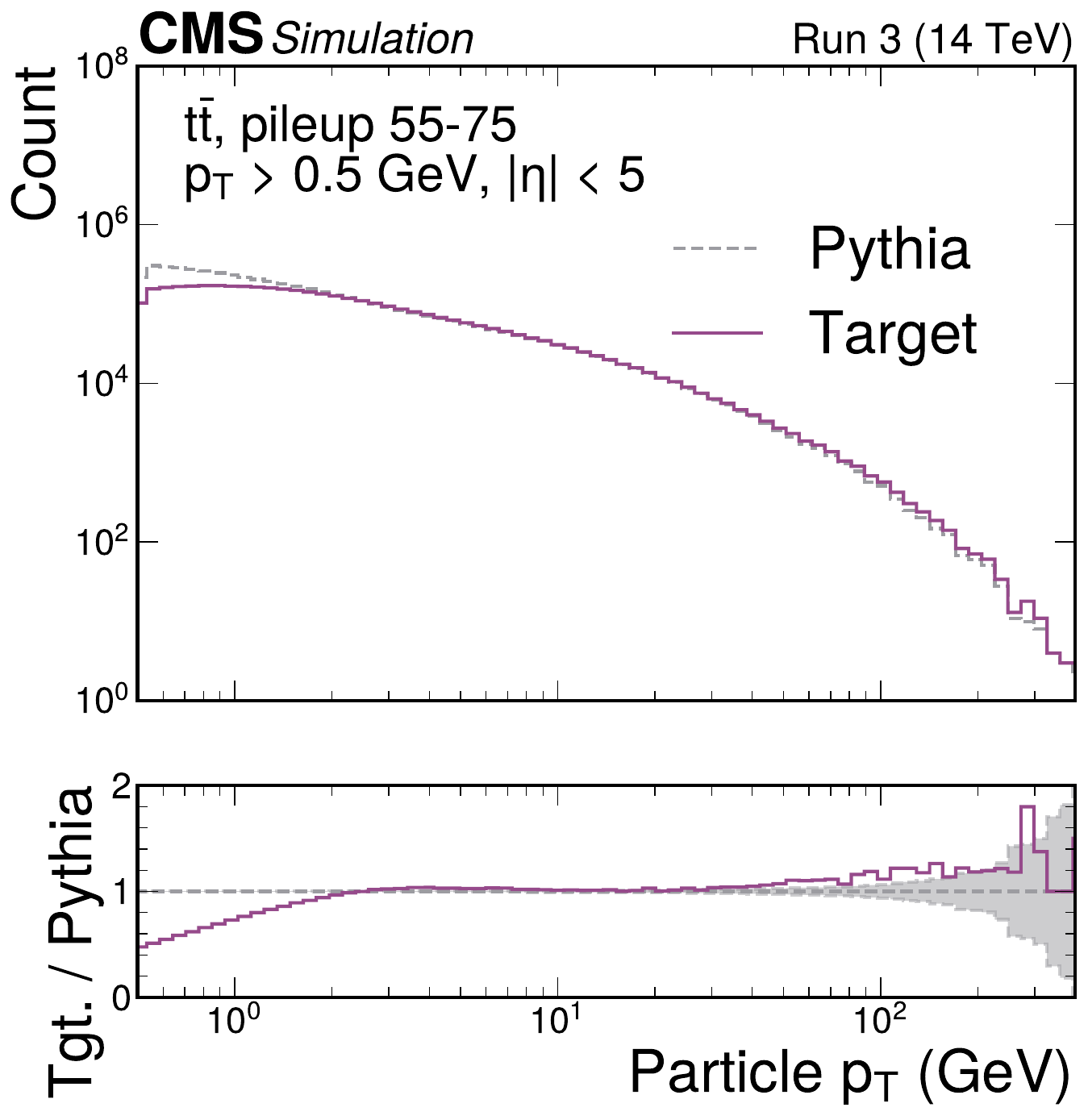}
    \includegraphics[width=0.38\linewidth]{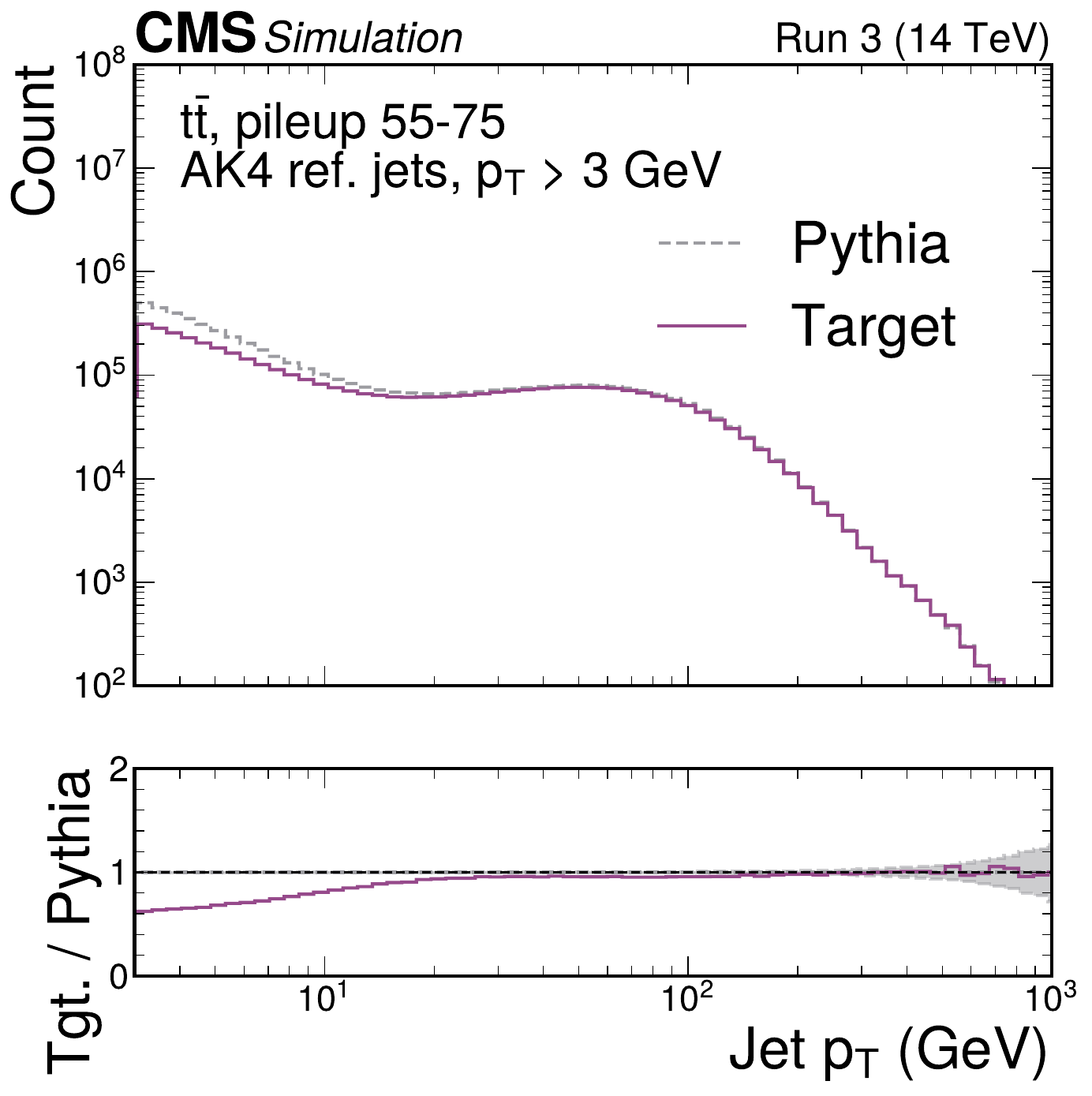}\\
    \hspace{0.062\linewidth}
    \includegraphics[width=0.38\linewidth]{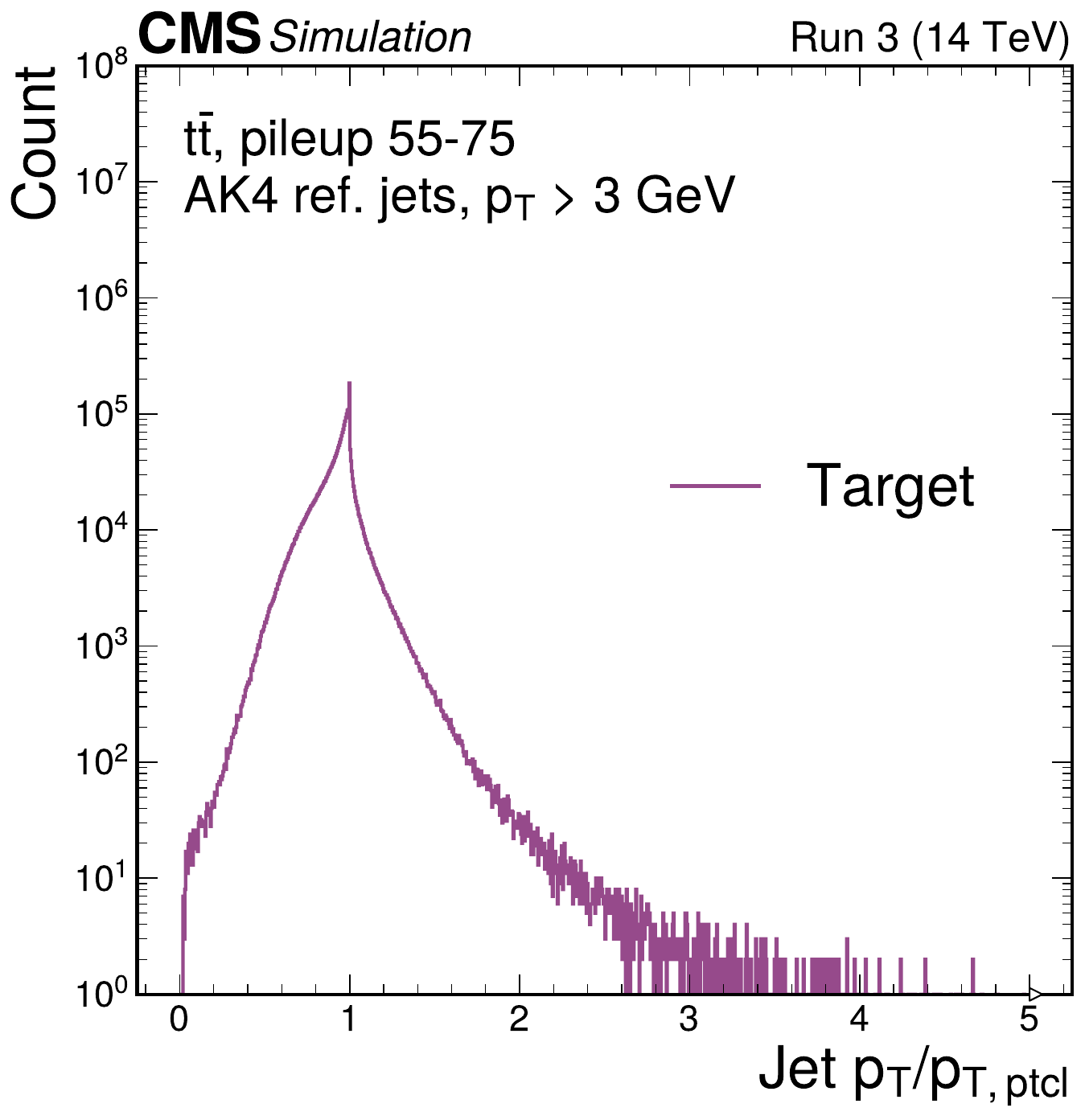}
    \includegraphics[width=0.45\linewidth]{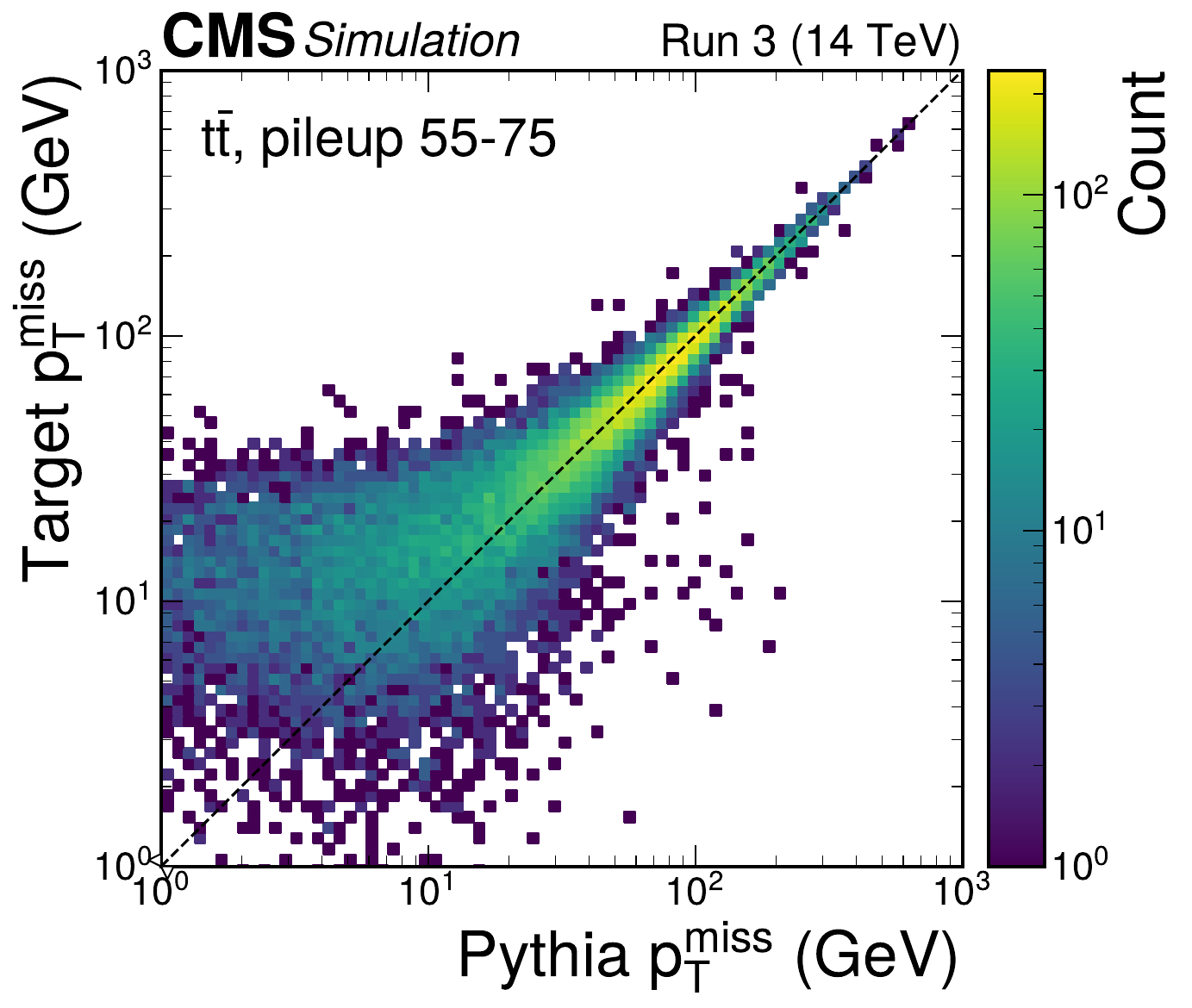}
    \caption{
        Validation of the MLPF particle-level target, after removing the contributions from the pileup particles, using \ttbar simulation with pileup.
        The \pt distribution of target particles and truth particles derived from \PYTHIA stable particles (upper left).
        The \pt distribution of jets clustered from each particle set (upper right).
        The jet response relative to particle-level jets (lower left).
        The target \ptmiss as a function of the \PYTHIA truth \ptmiss (lower right).
        The target particle set, after removing the contribution from pileup, generally aligns well with the pileup-free \PYTHIA truth.
    }
    \label{fig:target_validation}
\end{figure*}

We validate the MLPF target using inclusive \ttbar, \qcd multijet, and \Ztautau samples with pileup, confirming that the particle distributions generally align well with the truth after subtracting pileup contributions.
The \ttbar sample is shown in Fig.~\ref{fig:target_validation}.
The apparent discrepancy in the low-\pt portion of the particle and jet distributions is caused primarily by acceptance effects from the simulation and reconstruction, as very soft simulated photons are typically not propagated further through the detector simulation, or are not reconstructable because they leave no detectable hits.
The MLPF algorithm does not aim to correct for these effects.
The conclusions from the other samples for the target validation are similar.
The dependence of the physics performance on the precise target definition criteria can be further investigated in future work.

\subsection{Input features}
\label{sec:inp_features}
The input features for each track include the track \pt and its uncertainty, the $\eta$ and azimuthal angle $\phi$ (in radians) of the particle momentum and their uncertainties (both at the reference point, and extrapolated to the ECAL and HCAL), the Cartesian momentum components, the electric charge, the number of reconstructed hits in the track, and the coordinates of the closest approach of the track to the nominal beam line $(v_x, v_y, v_z)$.
The polar angle $\theta$, computed with respect to the beam axis, and $\lambda = \frac{\pi}{2}-\theta$~\cite{CMS:2014pgm}, along with their uncertainties, are also included.
Providing redundant angular variables such as $\eta$, $\theta$, and $\lambda$ allows the model to learn from representations with different numerical behavior across detector regions.
The model may use or disregard these redundant inputs, and we do not expect their inclusion to degrade the performance.
Finally, if the track contains a reference to a muon candidate, we also include the number of hits in the drift tubes and cathode strip chambers in the muon systems.

For calorimeter clusters, the input features include the Cartesian spatial coordinates ($x$, $y$, $z$), transverse energy (\ET), $\eta$, $\phi$, ECAL energy ($E_{\mathrm{ECAL}}$), HCAL energy ($E_{\mathrm{HCAL}}$), corrected energy and its associated uncertainty, and time and its uncertainty.
Additional features include the number of hits in the cluster and the cluster size, defined as the energy-weighted standard deviation of the hit positions in $x$, $y$, and $z$.
We also compute depth-dependent cluster shapes by evaluating the energy-weighted standard deviations in $\eta$ and $\phi$ as functions of the calorimeter layer (hit depth).

\subsection{The MLPF algorithm}
\label{sec:mlpfalgo}

We design a transformer-based architecture using self-attention~\cite{vaswani2017attention} for PF reconstruction, where the model predicts all reconstructed particles simultaneously in one inference step, as opposed to sequential or iterative approaches.
The inputs to the model are the reconstructed tracks and calorimeter clusters in an event, each represented by a feature vector $x_i$, collectively forming the set of $N$ PF elements $X$.
The model outputs reconstructed particle candidates $\hat{y}_i$, collected into a set $\hat{Y}$, and is trained to match the target set $Y$ containing target particles $y_i$.

    {
        \tolerance=10000
        Each target particle $y_i$, is characterized by the following quantities: (1) its particle identification (PID), (2) its four-momentum, parameterized as $(\pt, \eta, \sin{\phi}, \cos{\phi}, E)$, and (3) a binary label $\alpha$ indicating whether the particle originates from a pileup vertex.
        The latter serves solely to define the pileup-free target and is not used during the training; actual pileup mitigation is performed using the PUPPI algorithm.
        The azimuthal angle $\phi$ is parameterized through $\sin{\phi}$ and $\cos{\phi}$ to preserve the quadrant information.
        It may be beneficial to investigate adding position-encoded features based on $\eta$ and $\phi$ to improve spatial disambiguation in the model.
        \par
    }

Because the input and target sets generally have a different number of elements, we define a per-particle loss function by first assigning each target particle to exactly one unique input element, referred to as its \textit{primary element}.
This assignment is made from the set of target particles to the set of input elements (tracks or clusters), and is guided by physics-informed heuristics: charged particles are assigned to their originating tracks, and neutral particles to the calorimeter cluster where they deposit most of their energy.
For the case in which multiple particles contribute to a single cluster, we merge their energies and retain only a single particle type in the target, corresponding to the particle that deposited the largest fraction of the energy.
Although the underlying relationship between simulation-level particles and reconstruction-level inputs is inherently many-to-many, we empirically find that enforcing a one-to-one assignment from each target particle to a single input element enables a stable and physically meaningful loss function.
This formulation also naturally supports events with a variable number of particles.

The per-particle loss function follows the MLPF approach described in Refs.~\cite{Pata:2021oez,Pata:2022wam,Mokhtar:2023fzl}, and consists of three terms: a binary classification loss $L_\text{cls-binary}$ scaled by a factor of $\beta = 10$, a multiclass PID classification loss $L_\text{cls-PID}$, and a regression loss for the particle four-momentum $L_\text{reg}$.
The total event reconstruction loss is computed as a sum over all input tracks and clusters in an event:
\ifthenelse{\boolean{cms@external}}
{
    \begin{equation}
        \label{eq_loss}
        \begin{aligned}
            L(Y,\hat{Y}) & = \sum_{i} \Bigl[\beta \, L_\text{cls-binary}(y_i, \hat{y}_i) + L_\text{cls-PID}(y_i, \hat{y}_i) \\
                         & + L_\text{reg}(y_i, \hat{y}_i)\Bigr].
        \end{aligned}
    \end{equation}
}
{
    \begin{equation}
        \label{eq_loss}
        L(Y,\hat{Y}) = \sum_{i} \left[\beta L_\text{cls-binary}(y_i, \hat{y}_i) + L_\text{cls-PID}(y_i, \hat{y}_i) + L_\text{reg}(y_i, \hat{y}_i)\right].
    \end{equation}
}

{\tolerance=1200
    The $L_\text{cls-binary}$ term quantifies whether the corresponding track or cluster of the predicted particle is the primary element for a given target particle and is implemented using binary cross entropy, defined as the negative log-likelihood for a Bernoulli-distributed target.
    The $L_\text{cls-PID}$ term quantifies whether the predicted particle is classified correctly among particle types (\eg, electrons vs. photons) and is computed using focal loss~\cite{lin2017focal} to address class imbalance by emphasizing harder-to-classify examples.
    The $L_\text{reg}$ term quantifies how well the predicted particle's four-momentum components are estimated.
    To efficiently parameterize the \pt and $E$ losses, we transform the targets as:
    \begin{equation}
        \pt^\prime = \log\left( \frac{\pt^\text{target}}{\pt^\text{primary}} \right), \quad
        E^\prime = \log\left( \frac{E^\text{target}}{E^\text{primary}} \right),
    \end{equation}
    where $\pt^\text{primary}$ and $E^\text{primary}$ refer to the $\pt$ and $E$ of the associated primary input element (track or cluster).
    The model is trained to predict these transformed quantities, and the regression loss is computed using the mean squared error (MSE) between the predicted and target values:
    \ifthenelse{\boolean{cms@external}}
    {\begin{equation}
        \begin{aligned}
                L_\text{reg}(y, \hat y)  =& \gamma \,\mathrm{MSE}(\pt^\prime, \hat{p}_\mathrm{T}^\prime) + \gamma\, \mathrm{MSE}(E^\prime, \hat{E}^\prime)  \\
                                        &+ \delta\,\mathrm{MSE}(\eta, \hat{\eta}) + \delta\,\mathrm{MSE}(\sin\phi, \sin\hat\phi) \\
                                        &+ \delta\,\mathrm{MSE}(\cos\phi, \cos\hat\phi),
        \end{aligned}
    \end{equation}
    }
    {
        \begin{equation}
            L_\text{reg}(y, \hat y) = \gamma\, \mathrm{MSE}(\pt^\prime, \hat{p}_\mathrm{T}^\prime) + \gamma\, \mathrm{MSE}(E^\prime, \hat{E}^\prime) + \delta\, \mathrm{MSE}(\eta, \hat{\eta}) + \delta\, \mathrm{MSE}(\sin\phi, \sin\hat\phi) + \delta\, \mathrm{MSE}(\cos\phi, \cos\hat\phi),
        \end{equation}
    }
    where $\gamma$ and $\delta$ are weighting factors for the \pt and energy regression terms, and the angular regression terms, respectively.
    These transformed quantities lie within a more bounded range and are dimensionless, improving numerical stability in the optimization~\cite{holmberg2022jet}.
    \par}

However, this approach gives approximately equal weights to particles with different energy scales, potentially under-emphasizing high-\pt contributions.
To address this, we apply a per-particle \pt-dependent weighting factor $\gamma=\sqrt{\pt^{\text{target}}}$ to the \pt and energy regression terms to emphasize particles in the high-\pt tails.
We have observed that this weight term improves the jet energy scale and resolution performance over the baseline.
In addition, we apply the weighting factor $\beta=0.01$ to ensure that the three loss components are of comparable size during training.
All three loss terms are observed to converge smoothly, indicating that no single component dominates the total loss.

The transformer architecture learns pairwise correlations between all track and cluster pairs via \emph{scaled dot-product attention} (SDPA)~\cite{vaswani2017attention}.
In this mechanism, queries $Q$ and keys $K$ are $d_k$-dimensional learned projections of the input features, packed into $N{\times}d_k$ matrices, and the attention matrix is given by
\begin{equation}
    A = \mathrm{softmax}\!\left(\frac{QK^\top}{\sqrt{d_k}}\right),
\end{equation}
where the softmax operation is applied to normalize the attention weights such that they are positive and sum to unity for each query.
The element $A_{ij}$ represents the learned attention weight placed on the ``key'' input $x_j$ when updating the features of the ``query'' input $x_i$.
Although SDPA provides expressive pairwise interactions, it naively implies quadratic scaling in the number of inputs $N=\lvert X\rvert$ for both memory and computation time.
To mitigate the quadratic cost, we use \FlashAttention~\cite{dao2022flashattention}, an optimized attention kernel that achieves near-optimal GPU performance while preserving exact attention values, providing significant speedup and memory reduction.
The \FlashAttention kernel circumvents the need to materialize the full attention matrix by fusing the attention computation with the softmax operation and performing the computation in small tiles directly in fast on-chip memory, which significantly reduces memory usage and enables faster computation.
It is generally available pre-compiled for GPUs released after 2020, and can be compiled by hand for older GPUs (\eg, Turing).
However, CPUs currently do not support \FlashAttention because of the underlying hardware differences, but a fallback to the naive attention formulation via matrix multiplication is possible.

\begin{figure}[htbp]
    \centering
    \ifthenelse{\boolean{cms@external}}
    {
        \includegraphics[width=\columnwidth]{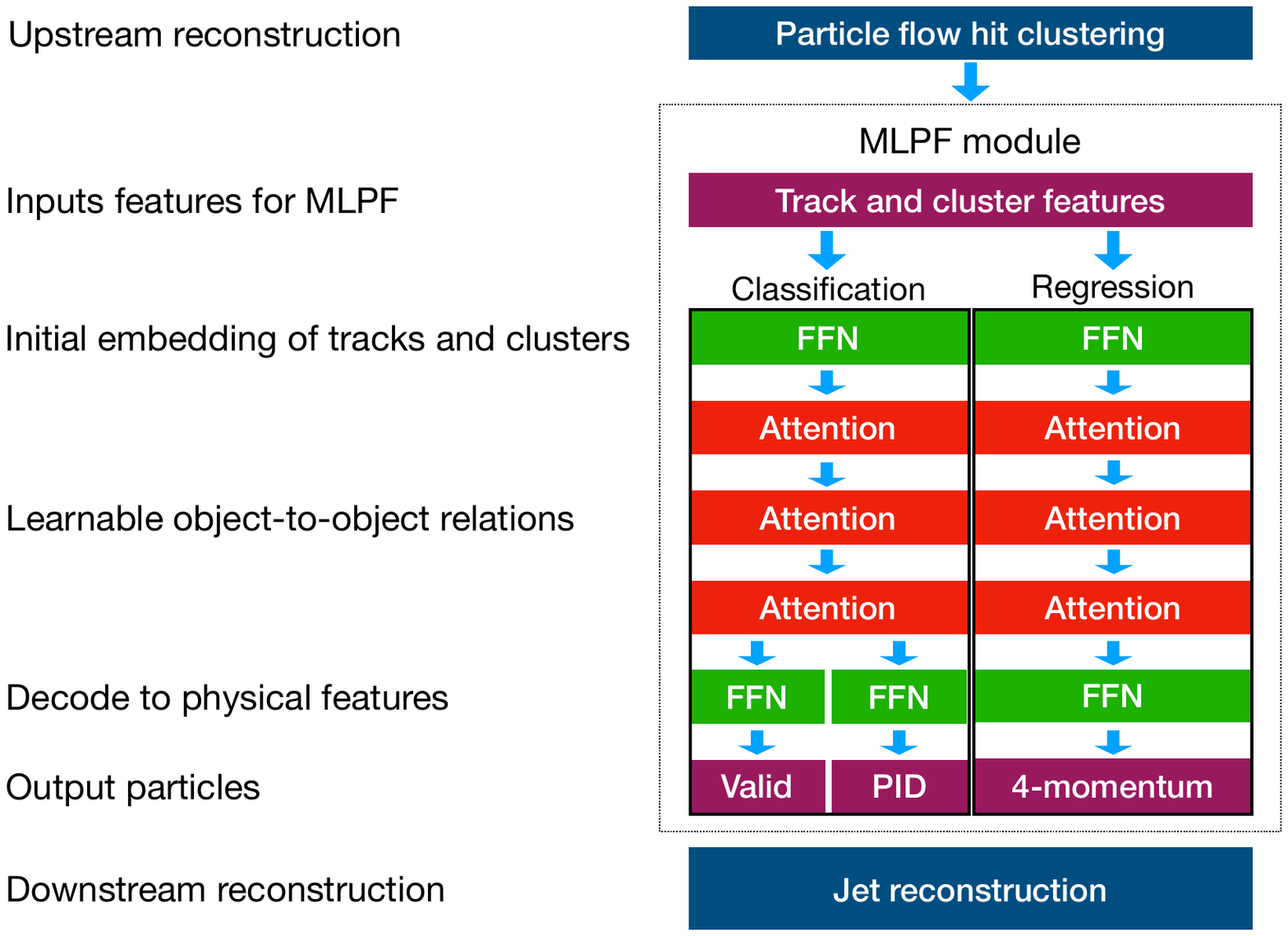}
    }
    {
        \includegraphics[width=0.8\linewidth]{Figure_002.pdf}
    }
    \caption{
        The neural network structure of the model.
        Input and output features are visualized in purple.
        Pointwise FFNs are visualized in green, and layers with full-event correlations in red.
        The last dimension contains a binary output to predict the presence of a particle (Valid), multi-class outputs for the PID, and the predicted momentum values.
    }
    \label{fig:model_structure}
\end{figure}

The architecture is illustrated in Fig.~\ref{fig:model_structure}, and is largely similar to the previous version tested in CMS~\cite{Pata:2022wam}, apart from replacing the GNN layers with transformer layers accelerated using \FlashAttention.
The input tracks and clusters are each embedded into a 256-dimensional latent space using simple feedforward networks (FFNs), allowing the model to jointly attend to both input types.
The embedded features are processed by two parallel transformer encoders: one for classification and one for regression, each comprising three transformer blocks with a dimension of $d_{\text{model}}=256$, comprising $N_{\text{heads}}=16$ heads with a head dimension $d_{\text{head}}=16$.
This allows the model to learn high-level correlations across the event, similarly to the baseline PF block algorithm.
Rectified linear unit activations are applied after each linear layer, and layer normalization is applied before each transformer block to improve convergence and reduce hyperparameter sensitivity~\cite{prelayernorm}.
Following the transformer encoders, six two-layer FFN heads produce the final outputs.
Two heads, fed by the classification encoder, perform the binary primary target particle classification and multi-class PID.
The remaining four heads, fed by the regression encoder, are dedicated to the regression tasks.
Finally, the encoded representations are decoded to predict particle candidates using the per-particle loss defined in Eq.~(\ref{eq_loss}).
At inference time, the set of predicted particles is constructed by selecting the candidates with a binary classification score above 0.5, and assigning them the PID with the largest probability score and four-momentum from the corresponding heads.
While this paper focuses on full event inputs, we note that it may be interesting to investigate partitioning the event before the correlation-building layers to improve generalization while keeping edge effects under control~\cite{Kakati:2024dun}.

\begin{figure*}[htbp]
    \centering
    \includegraphics[width=0.32\linewidth]{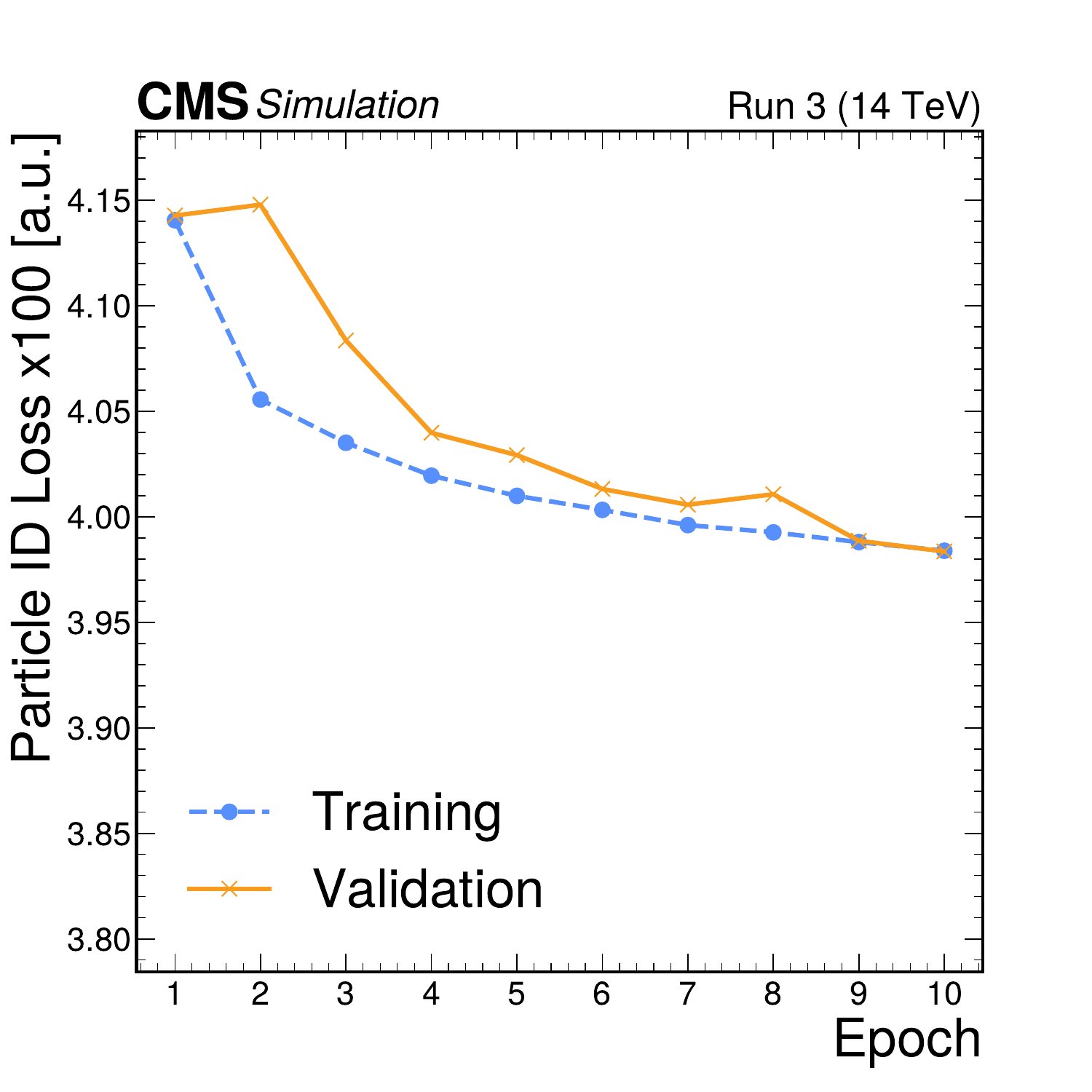}
    \includegraphics[width=0.32\linewidth]{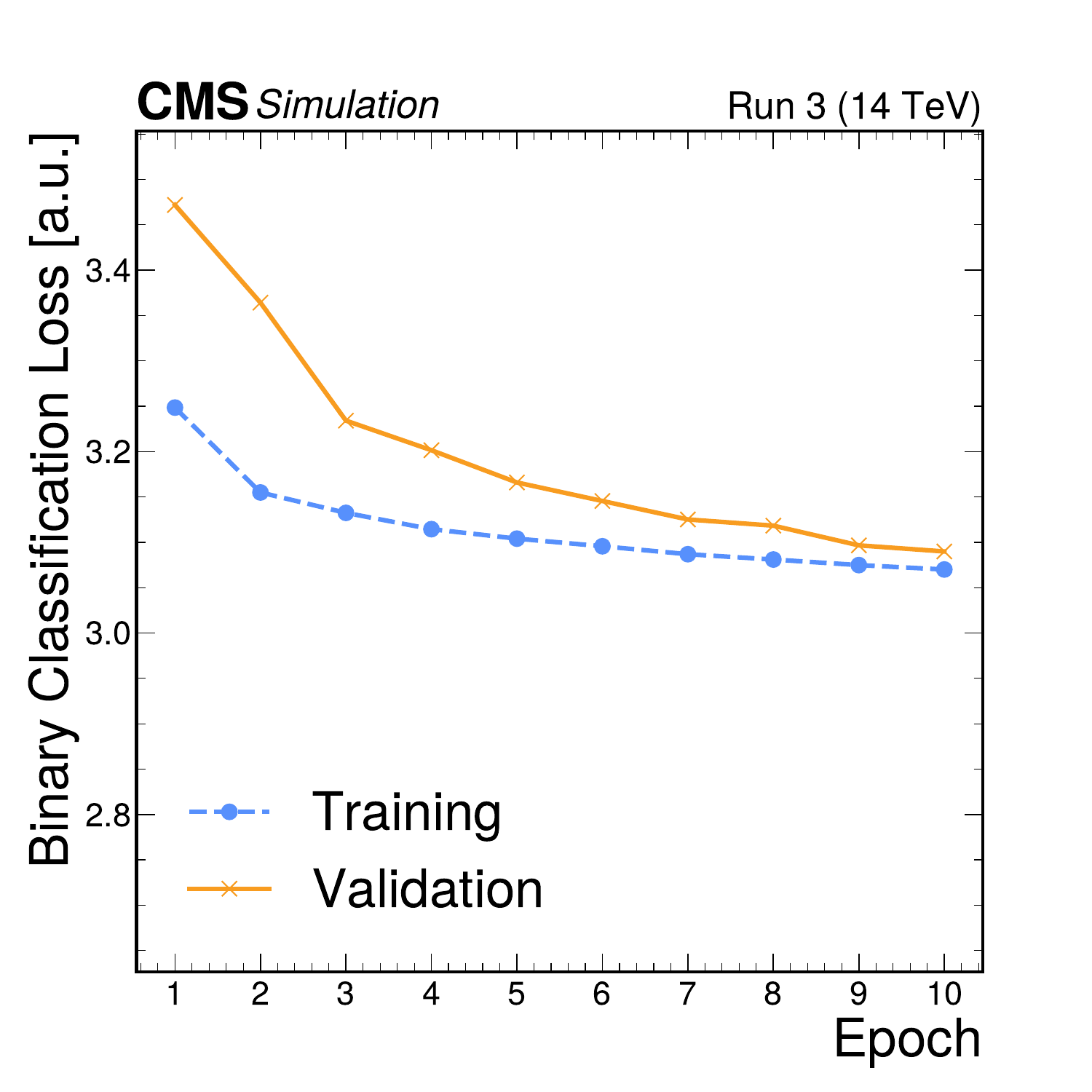}
    \includegraphics[width=0.32\linewidth]{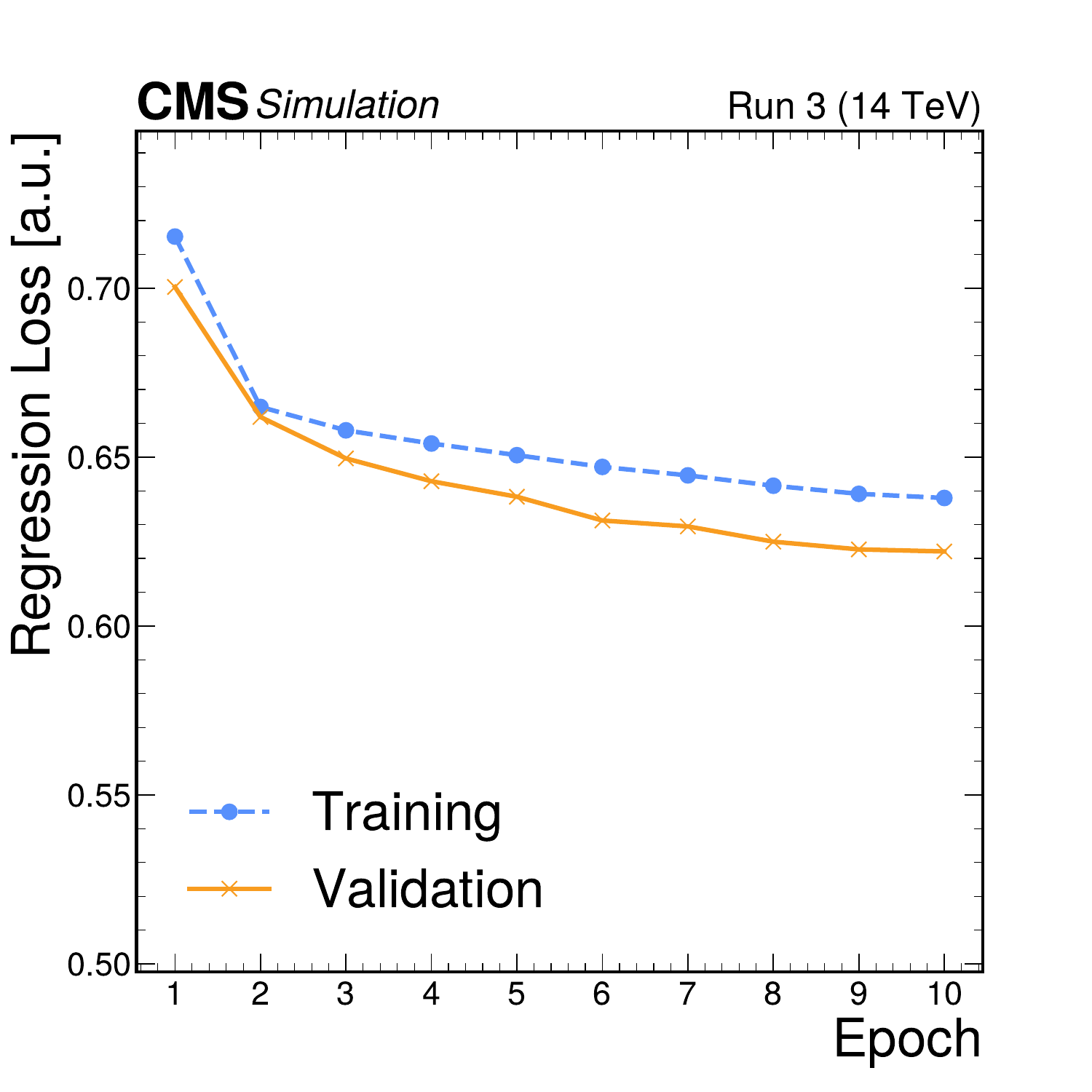}
    \caption{
        The PID, binary classification, and regression losses for the training and validation data sets, as a function of the training epoch.
        We observe both the training and validation losses to be smoothly converging.
        The training loss is evaluated with a partially-trained model throughout the training epoch on the training events (80\%), while the validation loss is computed with the model weights at the end of the epoch on the validation events (20\%).
        The losses are measured in arbitrary units, averaged across the events.
    }
    \label{fig:model_loss}
\end{figure*}

The model is trained end-to-end using an open-source software repository~\cite{joosep_pata_2025_17519540} based on \PYTORCH~\cite{pytorch} (unmodified version 2.7.1 with ROCM 6.2.4) on eight AMD MI250X GPUs for ten epochs, requiring approximately six hours per epoch.
AMD GPUs were used due to their availability on the LUMI HPC, while the training has also been confirmed to work equivalently on Nvidia GPUs.
All loss components---binary classification, PID, and regression---converge smoothly with no signs of overfitting, as shown in Fig.~\ref{fig:model_loss}, and we currently see no benefit from additional training or from additional self-attention layers in the model.
The training time can be reduced by distributing the workload across multiple GPUs using a simple data-parallel setup, along with the appropriate scaling of the optimizer parameters.
We use the LAMB optimizer~\cite{You2019LargeBO} suitable for large-batch optimization, with an initial learning rate of 0.002, decaying to 0.0002 at the end of the training with a cosine annealing schedule.
The current model contains approximately 5.1 million trainable parameters, which is of comparable size to the unified particle transformer architecture used for flavor jet tagging in CMS~\cite{CMS-DP-2024-066,CMS-DP-2025-081}.
While the total number of training events is currently limited by data set disk size and available computing resources, it is important to note that a single event with pileup corresponds to roughly 5,000 input elements, providing substantial training information per event.
Empirically, we observe that increasing the data set size leads to improvements in performance, particularly in the distribution tails.

\section{Physics performance in simulation}
\label{sec:perf_sim}

The MLPF algorithm is validated directly using the offline software on dedicated testing data sets, by replacing the standard PF reconstruction with MLPF, using the integration described in Section~\ref{sec:cmssw}.
We compare the performance of PF and MLPF in \qcd multijet and \ttbar samples in terms of particle-level reconstruction, and the performance of jets and \ptmiss.

\subsection{Particles}
\label{sec:perf_sim_particles}

\begin{figure*}[htbp]
    \centering
    \includegraphics[width=0.49\linewidth]{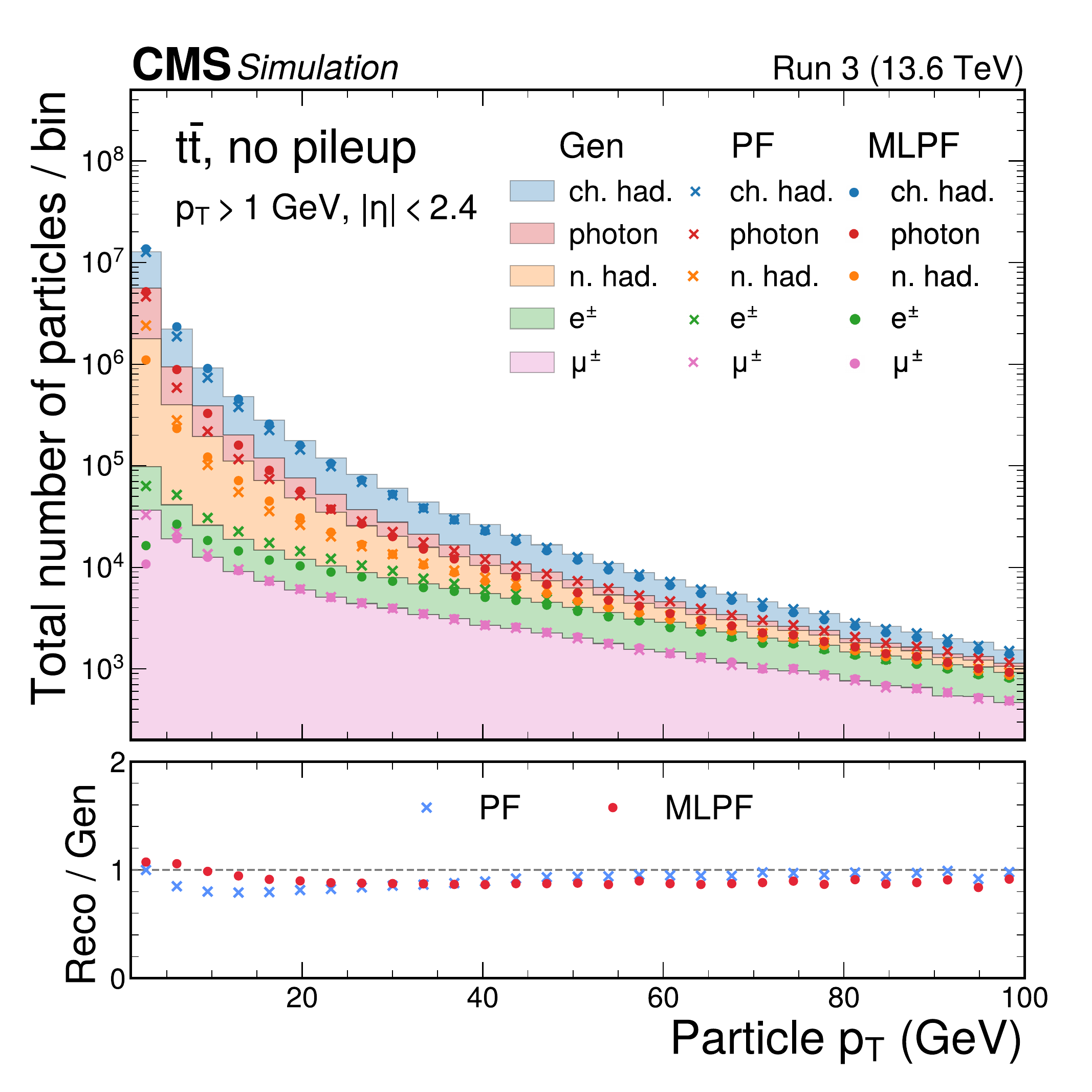}
    \includegraphics[width=0.49\linewidth]{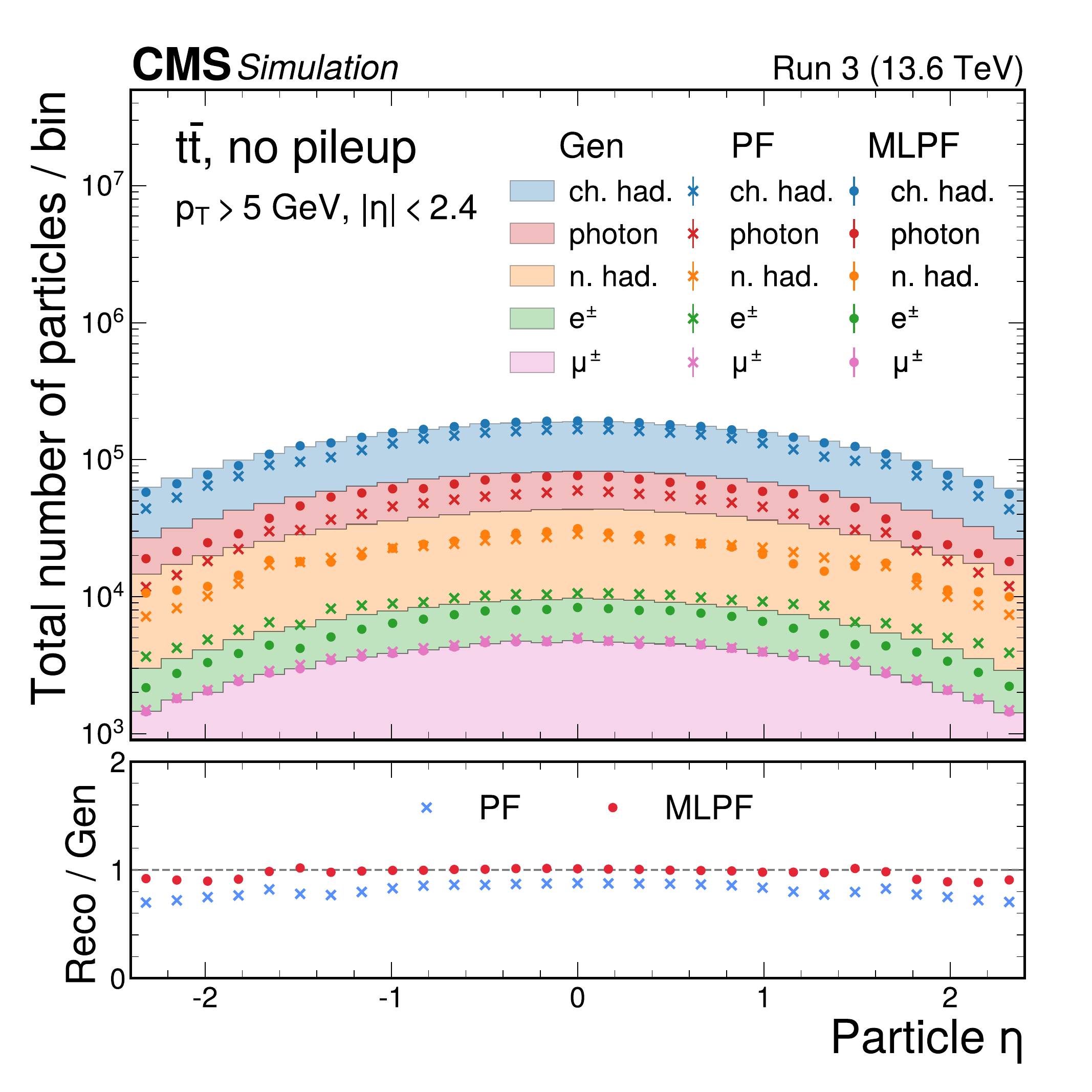}
    \caption{
        The particle \pt (left) and $\eta$ (right) distributions split by PID for MC particles (Gen), and PF and MLPF candidates in \ttbar events.
        In the forward region, the PF and MLPF algorithms do not attempt to identify particles, but instead reconstruct the deposits as either electromagnetic or hadronic.
    }
    \label{fig:perf_particles_nano_stacked}
\end{figure*}

To characterize the reconstruction of individual particles, we first apply a $\pt>1\GeV$ selection to both generator-level \PYTHIA particles and PF or MLPF candidates.
This ensures that we restrict the comparison to particles energetic enough to produce detectable signals; as shown in Fig.~\ref{fig:target_validation}, generator-level particles below this threshold seldom appear in the target set.
The single particle distributions in \pt and $\eta$, split by individual particle type, in \ttbar samples is shown in Fig.~\ref{fig:perf_particles_nano_stacked}.
We restrict the distributions to the central region ($\abs{\eta}<2.4$), where the detector is more granular and the reconstruction is more complex because of the overlapping tracker and calorimeters.
Moreover, particle identification is more reliable in this region, making it meaningful to show separate distributions for each particle type.
For the $\eta$ distribution, we increase the \pt threshold to 5\GeV to suppress the low-\pt region, where discrepancies between generator-level and reconstructed particles are more pronounced.
Overall, we observe a comparable level of performance between MLPF and PF across most particle types.
The stacked multiplicity plot is intended to provide a qualitative illustration of the global event composition rather than a detailed assessment of the reconstruction performance for individual particle types.

{\tolerance=1000
To measure the particle-level efficiency and misidentification rate, generator-level \PYTHIA particles are matched to PF or MLPF candidates within a cone of $\Delta R < 0.15$, where $\mbox{$\Delta R = \sqrt{(\Delta\eta)^2 + (\Delta\phi)^2}$}$ denotes the angular distance in the $(\eta, \phi)$ plane.
Matches are assigned uniquely by iteratively pairing the closest unmatched generator-reconstructed particle pair in $(\eta,\phi)$ space, ensuring that multiple matches do not occur.
We have explored variations of the matching cone size and found that the chosen $\Delta R$ threshold provides a reasonable compromise that is small enough to avoid detector-resolution effects dominating the matching, yet large enough to prevent excessive overlap that would push the particle matching into the jet-clustering regime.
The efficiency is defined as the fraction of generator-level particles that are matched to any PF or MLPF candidate, while the misidentification rate is defined as the fraction of PF or MLPF candidates that are not matched to any generator-level particle.
\par
}

\begin{figure*}[htbp]
    \centering
    \includegraphics[width=0.45\linewidth]{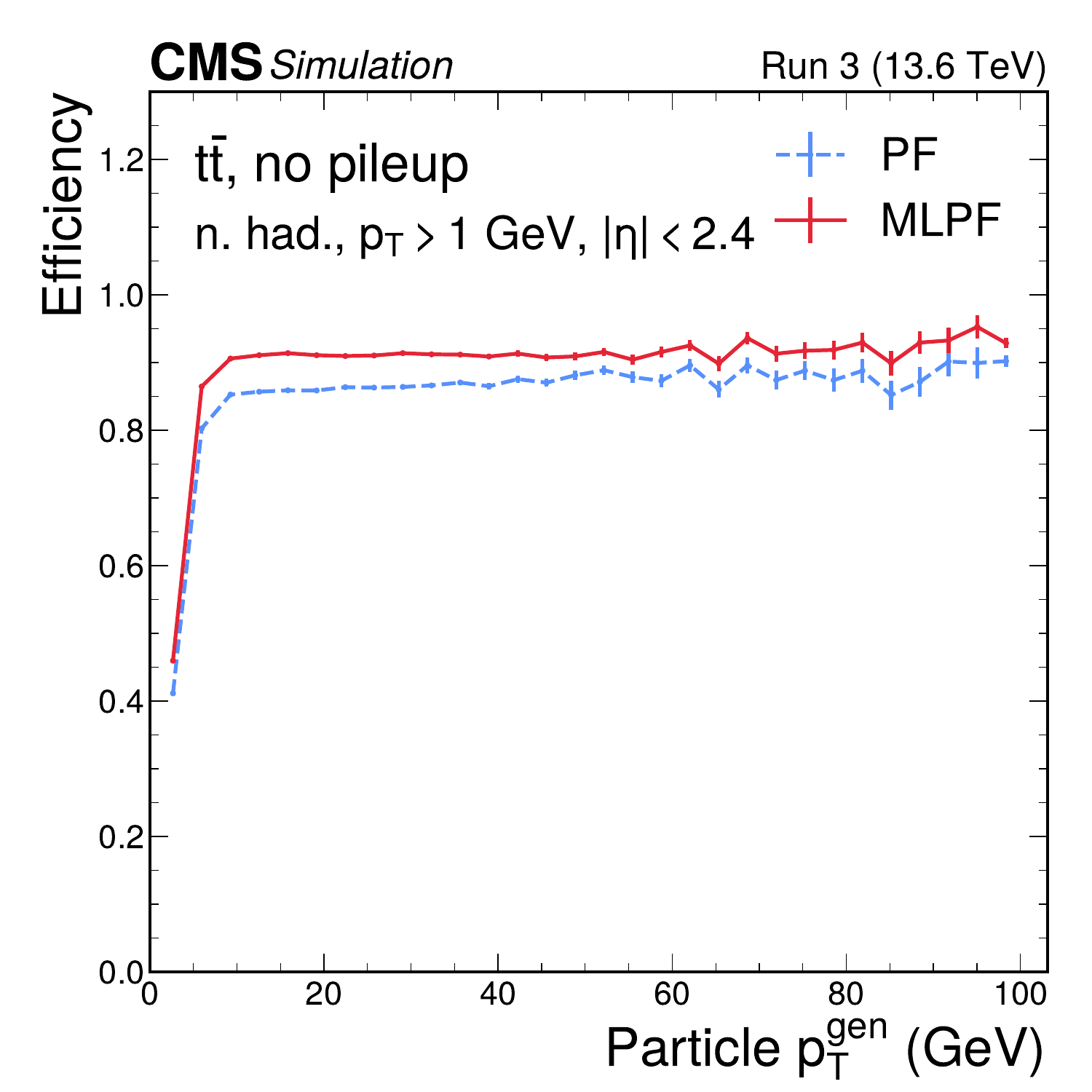}
    \includegraphics[width=0.45\linewidth]{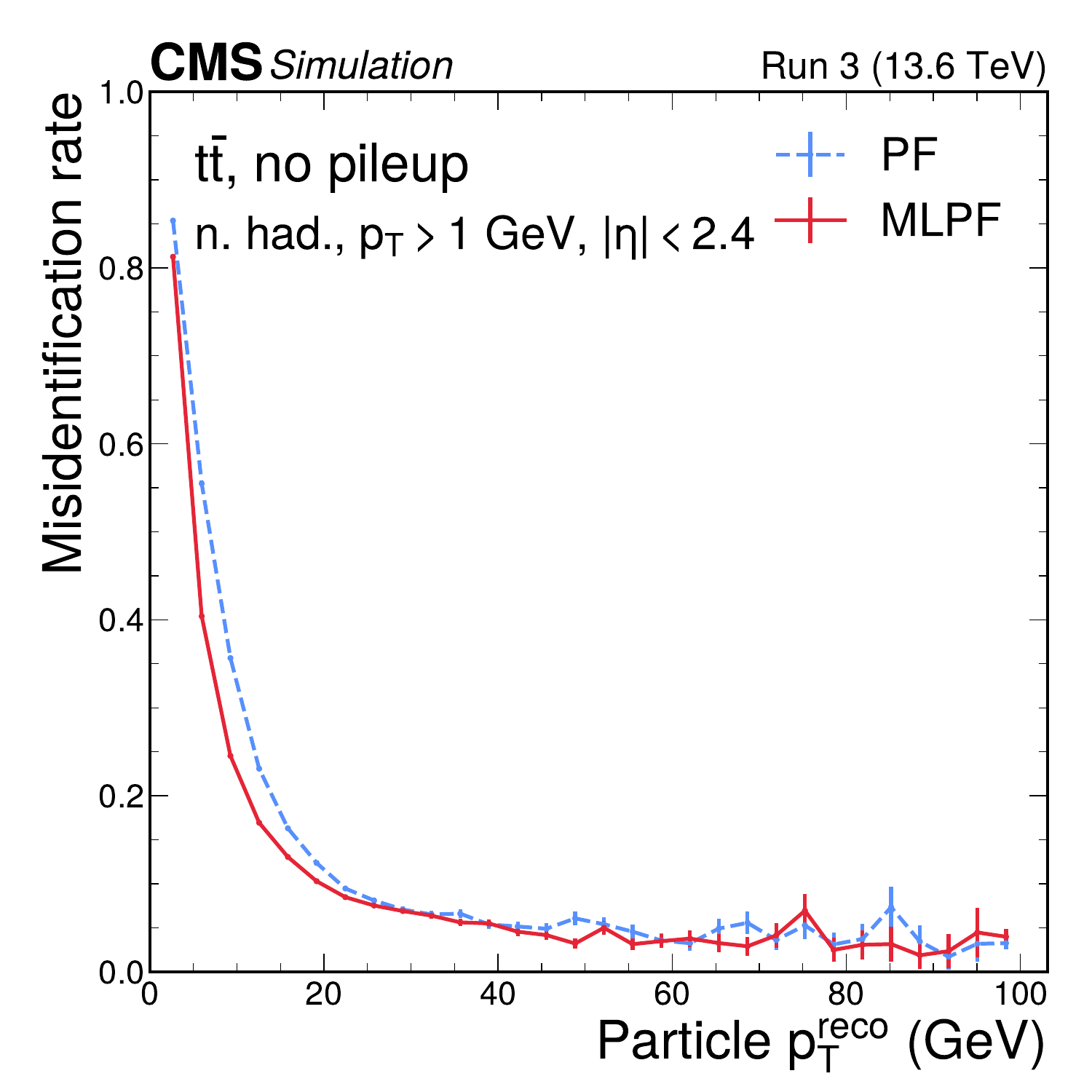}
    \hspace{0.025\linewidth}
    \caption{
        Neutral hadron efficiency (left) and misidentification rate (right) as functions of \pt for PF (blue) and MLPF (red) in \ttbar events based on $\Delta R$ matching to generator-level particles.
        The MLPF algorithm generally has higher efficiency at a lower misidentification rate.
        The vertical bars indicate the statistical uncertainties on the respective distributions.
    }
    \label{fig:perf_particles_nano_eff}
\end{figure*}

Figure~\ref{fig:perf_particles_nano_eff} shows the efficiency and misidentification rate for neutral hadrons in \ttbar samples.
No PID requirement is applied during the matching.
The MLPF algorithm achieves improved neutral hadron efficiency while maintaining the same misidentification rate as standard PF.
In addition, the energy regression of neutral particles is found to be consistent between the two algorithms.
For charged hadrons and photons, MLPF achieves higher efficiency with a slightly higher misidentification rate as a function of \pt.
For completeness, we note that MLPF exhibits a slightly lower efficiency and misidentification rate for electrons compared to PF, with a fraction of electrons being misreconstructed as charged hadrons, particularly at low-\pt.
For isolated high-\pt electrons, the performance of PF and MLPF is nearly identical.
For muons, the efficiency and misidentification rate are found to be similar between MLPF and PF.
Finally, we note that the dedicated electron and muon reconstruction sequences, which take PF candidates as input in the standard CMS reconstruction chain, have not yet been retuned for MLPF.
Consequently, a detailed study of the performance for isolated electrons and muons is left for future work.

The working point for particle classification in MLPF can be tuned to enhance the efficiency vs. misidentification rate for certain classes; however, in this work, we opt for a naive assignment to the PID with the largest probability score.
It is important to note that these metrics are sensitive to fiducial requirements, $\Delta R$ thresholds, and other factors, such as the ambiguity of matching in dense environments.

\subsection{Jets and \texorpdfstring{\ptmiss}{pTmiss}}
\label{sec:perf_sim_jets}

Jets are clustered using the anti-\kt algorithm with distance parameters $R=0.4$ (AK4) and $0.8$ (AK8), using either PF or MLPF candidates within the offline software.
Jet performance in simulation is evaluated by matching reconstructed jets to generator-level jets within a cone of $\Delta R < 0.2$.
For both PF and MLPF jets, the PUPPI algorithm is applied on top of the reconstructed particles to mitigate the effects of pileup before the jet clustering is performed.
Jet performance is quantified using the jet response, defined as the ratio of reconstructed jet \pt and generator-level jet \pt.

\begin{figure}[htbp!]
    \centering
    \ifthenelse{\boolean{cms@external}}
    {
        \includegraphics[width=\columnwidth]{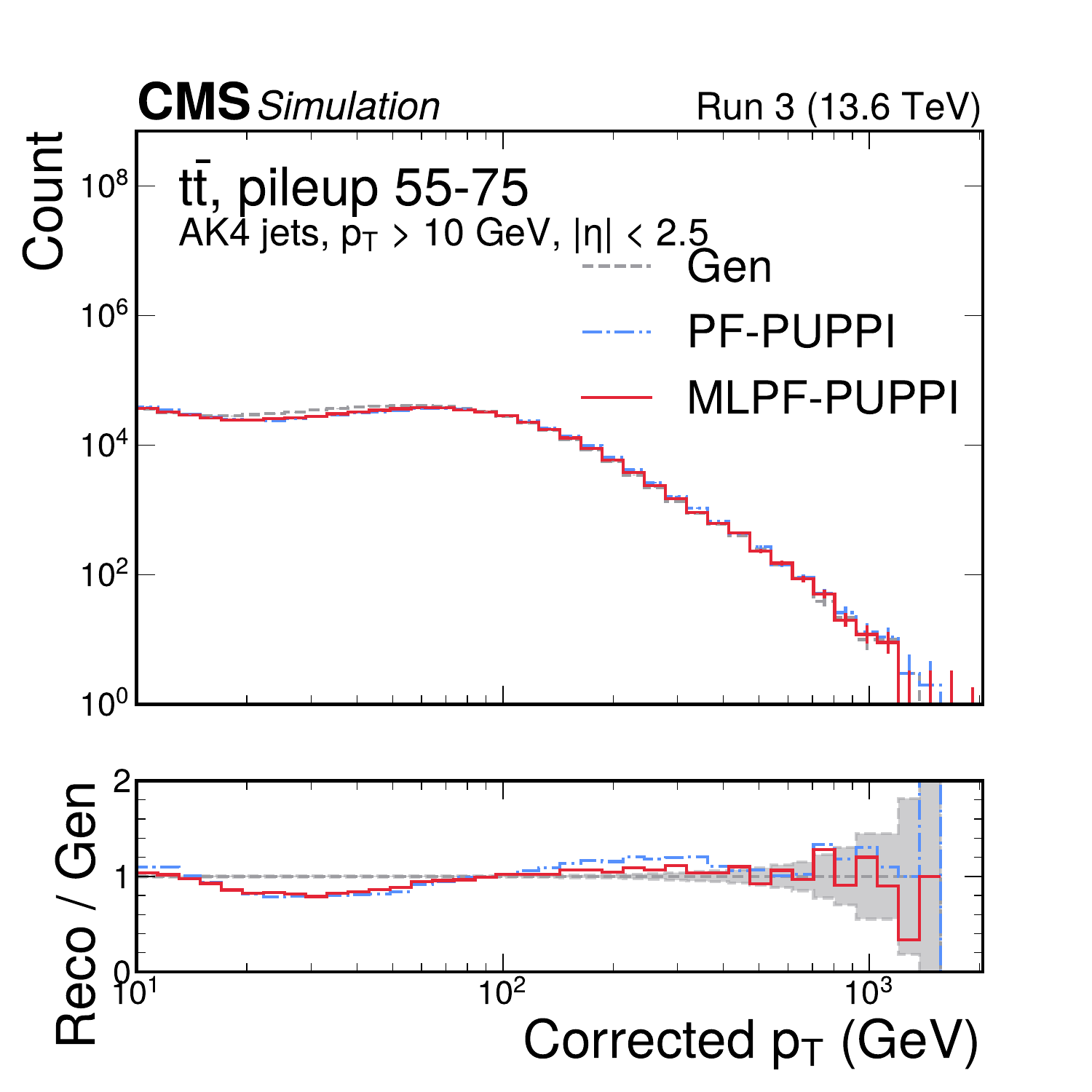}
        \includegraphics[width=\columnwidth]{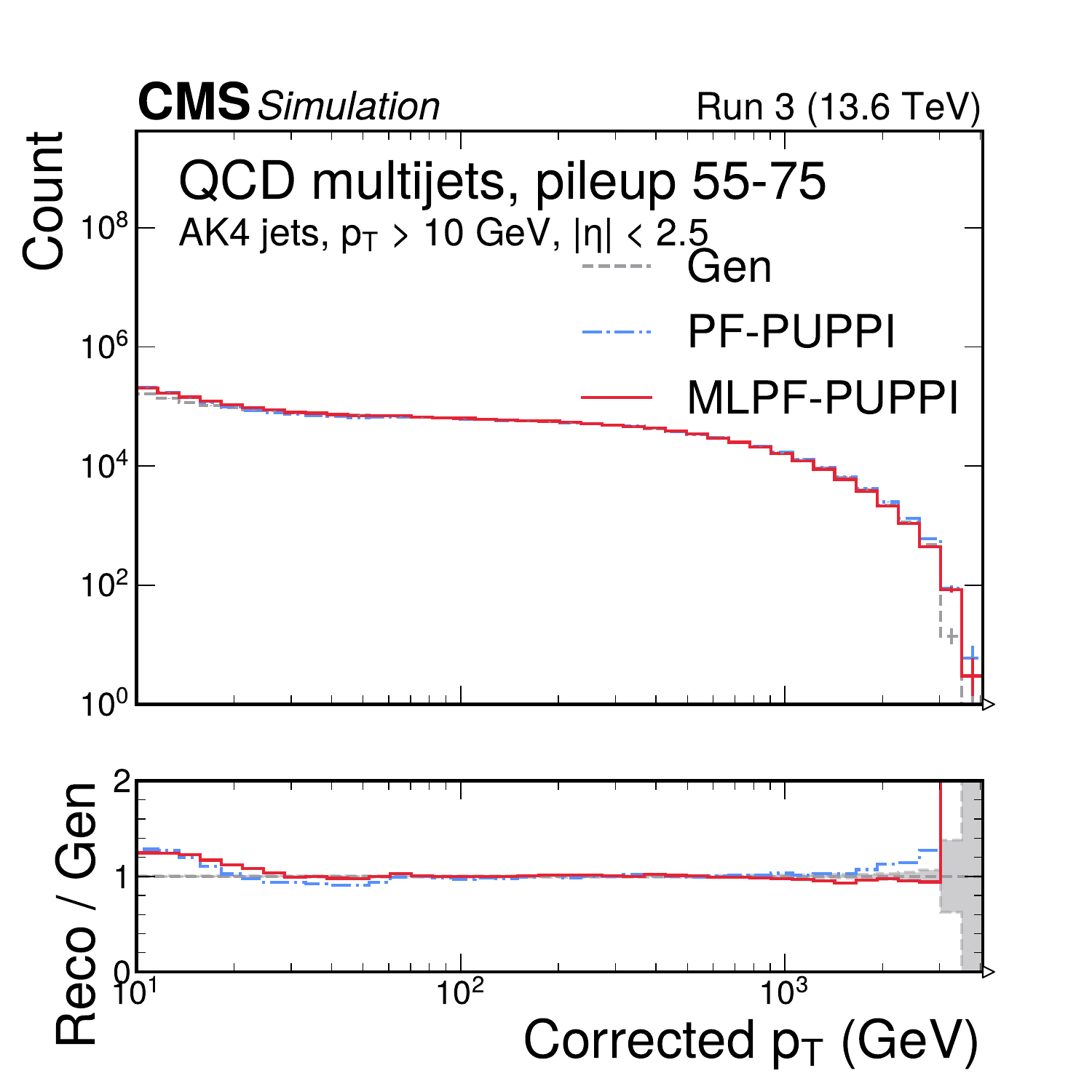}
    }
    {
        \includegraphics[width=0.47\linewidth]{Figure_006-a.pdf}
        \includegraphics[width=0.47\linewidth]{Figure_006-b.pdf}
    }
    \caption{
        The corrected jet \pt distribution in \ttbar (\cmsLeft) and \qcd multijet (\cmsRight) validation samples with 55--75 pileup interactions.
        MLPF is only trained to reconstruct particles, and is never explicitly optimized to reconstruct jets.
        The PUPPI algorithm is applied to both PF and MLPF jets to perform pileup subtraction.
        We show the jet \pt distribution after simulated response calibration.
        The shaded bands indicate the statistical uncertainties on Gen.
    }
    \label{fig:perf_jets_mc_pt}
\end{figure}

{\tolerance=1200
For jet performance, we focus on the region with $\abs{\eta}<2.5$, as this is the most relevant domain for PF studies due to the simultaneous use of the tracker and calorimeters.
Within this central region, we derive jet energy corrections using standard CMS jet energy calibration procedures to perform the simulated response calibration for both PF and MLPF.
Events are first selected by requiring the longitudinal distance between the reconstructed PV and the generator-level PV to be less than 0.2\unit{cm}.
The three most energetic generator-level jets before calibration are then used for the subsequent response evaluation.
Generator-level jets are matched to reconstructed jets using an iterative matching procedure, and the jet energy scale is calibrated by measuring the jet energy response in bins of reconstructed-jet $\eta$ and generator-level jet \pt.
The calibration is performed separately for jets clustered from PF and MLPF particle candidates.
For each bin, the median jet energy response is computed, and the correction is taken as the inverse of this median, which is linearly interpolated between the surrounding bin centers.
The corrected jet \pt distributions, shown in Fig.~\ref{fig:perf_jets_mc_pt}, are compatible between PF and MLPF.
Although the reconstructed jets are brought closer to the generator-level spectrum through the simulated response calibration procedure, perfect agreement is not expected.
Residual differences arise from standard reconstruction inefficiencies and from jet-identification requirements applied only at reconstruction level.
\par
}

\begin{figure*}[htbp]
    \centering
    \includegraphics[width=0.47\linewidth]{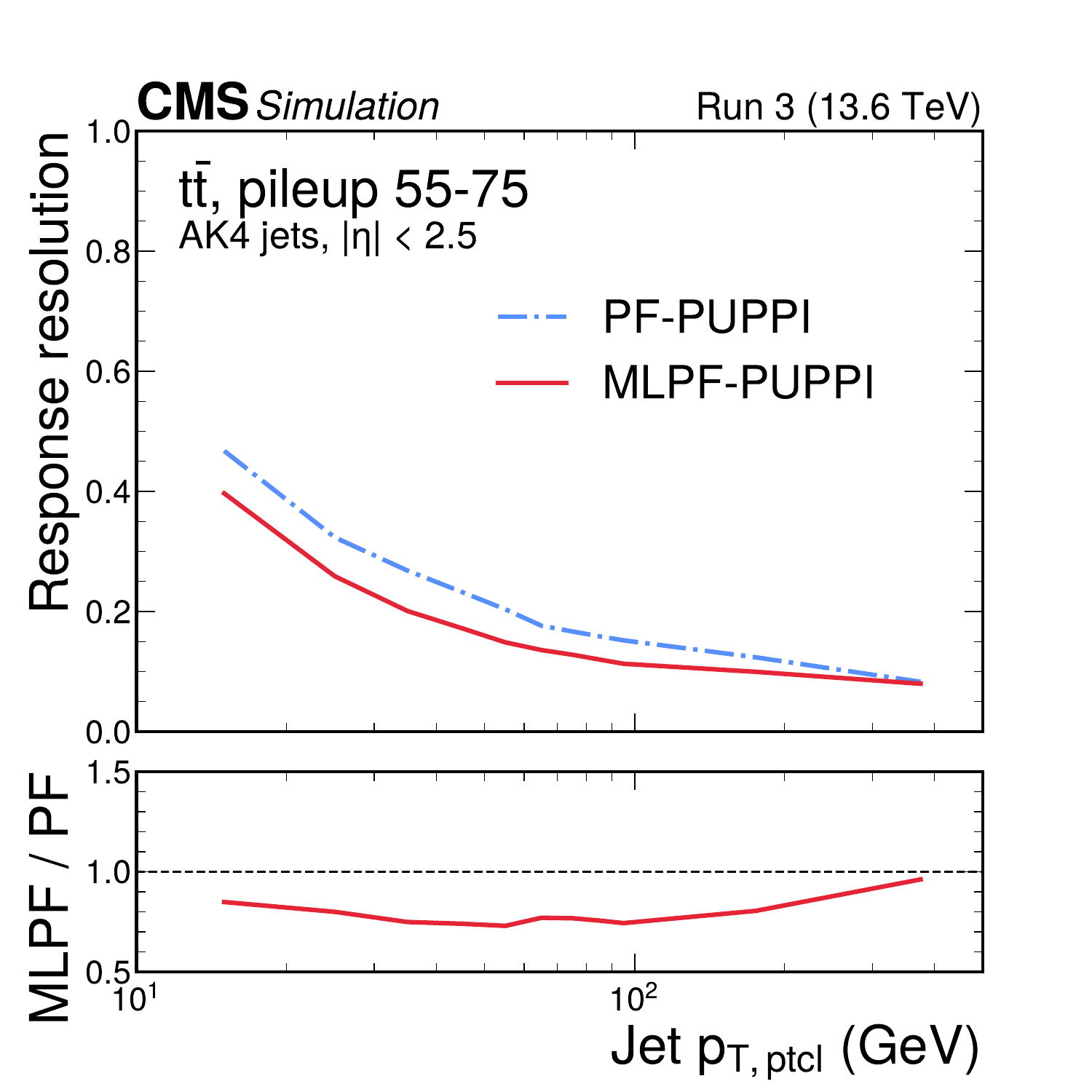}
    \includegraphics[width=0.47\linewidth]{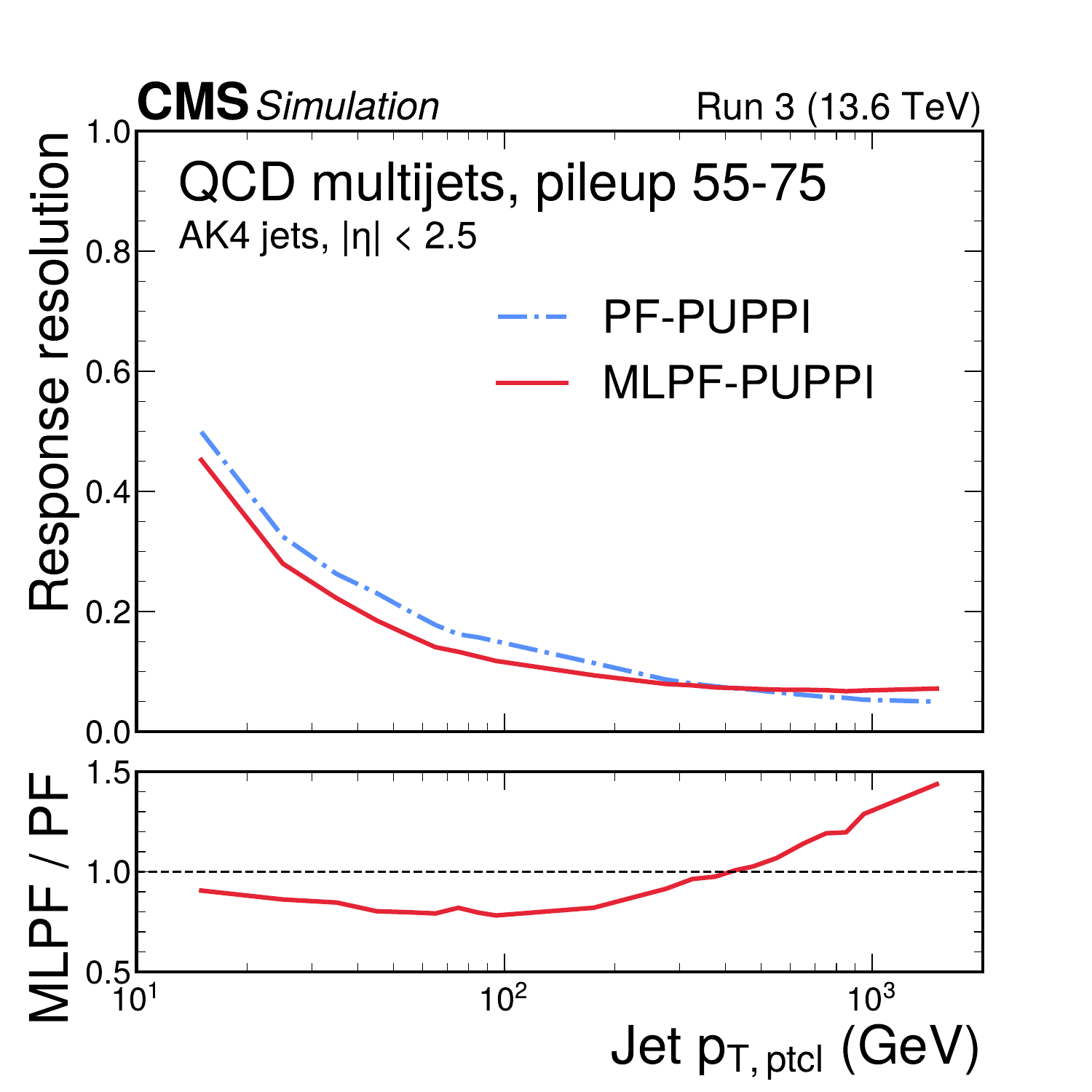}
    \caption{
        The jet resolution in \ttbar (left) and \qcd multijet (right) validation samples with 55--75 pileup interactions, after simulated response calibration.
        The MLPF algorithm is only trained to reconstruct particles, and is never explicitly optimized to reconstruct jets.
        The PUPPI algorithm is applied to both PF and MLPF jets to perform pileup subtraction.
        The jet resolution is parameterized by fitting a Gaussian distribution and dividing the best-fit standard deviation by the mean, showing the bin midpoints on the horizontal axis.
    }
    \label{fig:perf_jets_mc_reso}
\end{figure*}

\begin{figure*}[htbp]
    \centering
    \includegraphics[width=0.47\linewidth]{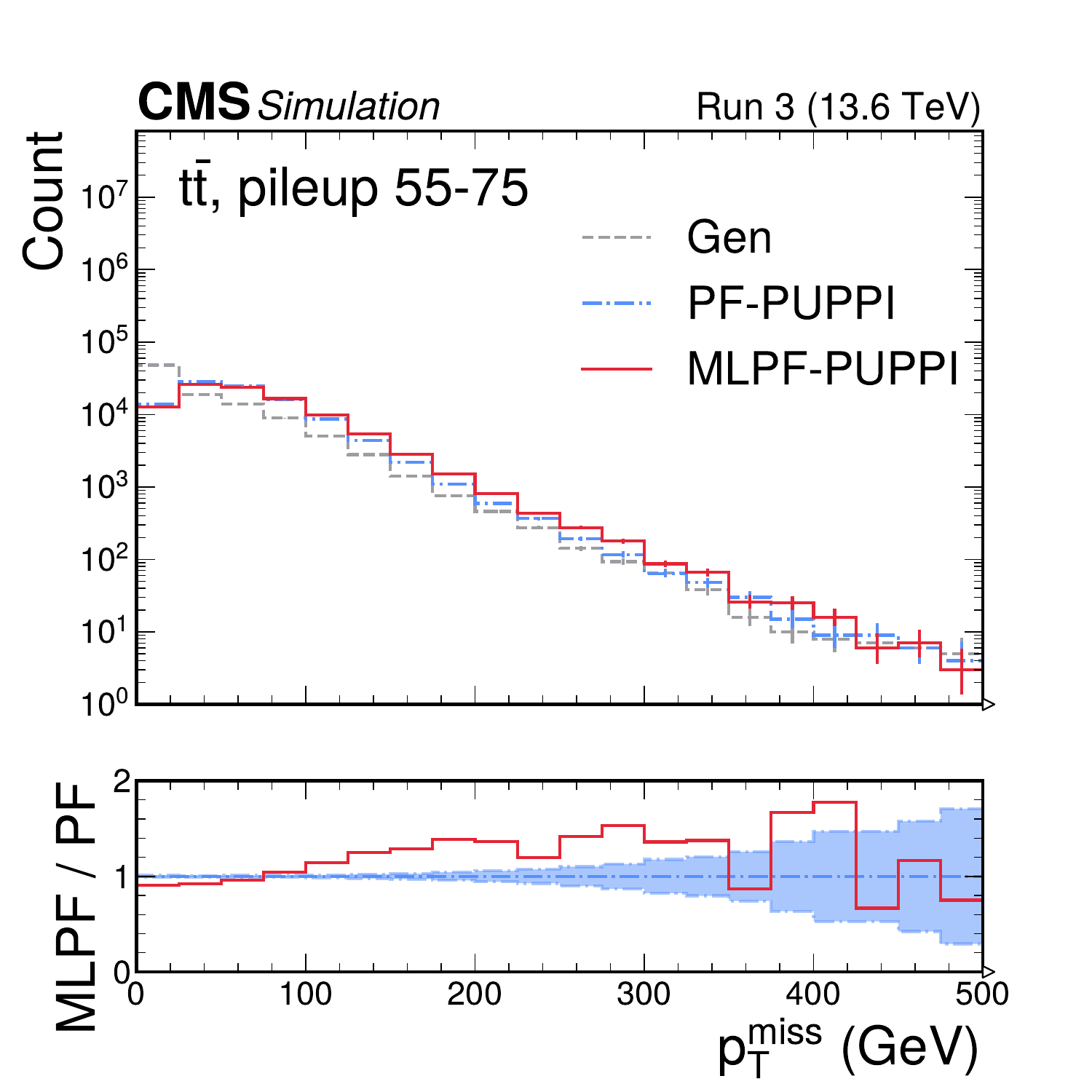}
    \includegraphics[width=0.47\linewidth]{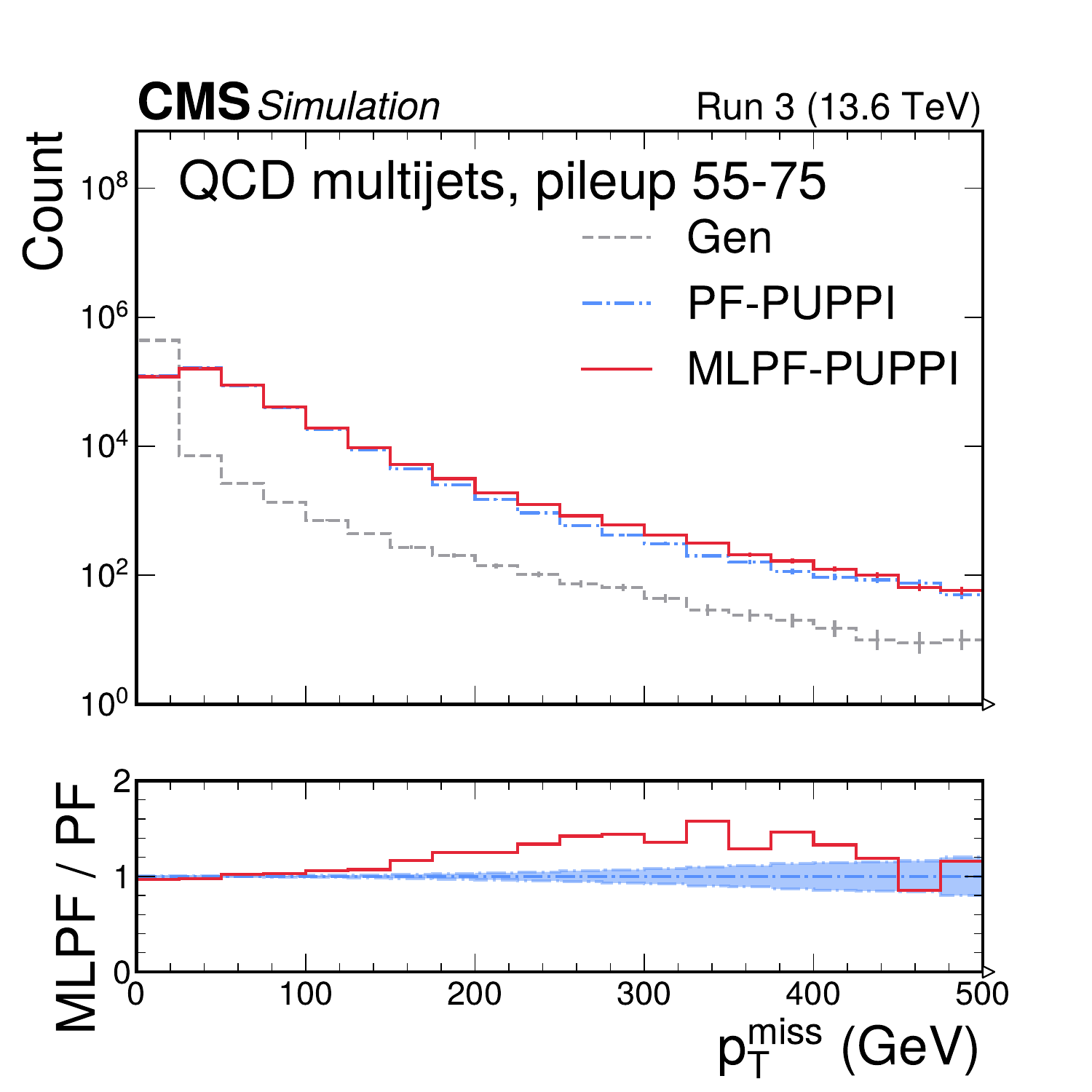}
    \caption{
        Uncalibrated \ptmiss in \ttbar (left) and \qcd multijet (right) samples with pileup from offline reconstruction using PF and MLPF.
        The MLPF algorithm is only trained to reconstruct particles, and is never explicitly optimized to reconstruct \ptmiss.
        The \ptmiss distributions from PF and MLPF are found to be generally consistent, with MLPF reconstructing a somewhat harder \ptmiss spectrum in the tails.
        The vertical bars indicate the statistical uncertainties on the respective distributions, and the shaded bands in the ratio panel indicate the statistical uncertainties on PF.
    }
    \label{fig:perf_met_nopu_cmssw}
\end{figure*}

As shown in Fig.~\ref{fig:perf_jets_mc_reso}, we observe a 10--20\% improvement in jet energy resolution with MLPF compared to standard PF for jets with \pt between 30--100\GeV.
This improvement is observed consistently in both \ttbar and \qcd multijet events, demonstrating that the conclusions are robust across samples with different underlying physics processes.
In the high-\pt region, the relative improvement of MLPF decreases, which we attribute to the limited number of high-\pt training samples.
Increasing the size of the training dataset has been observed to improve the tails of the MLPF distributions, supporting this interpretation.
Fine-tuning on dedicated high-\pt samples is therefore in principle possible and can be performed by analyses that specifically target the boosted regime.
Finally, it is important to note that the MLPF model is trained using a per-particle loss function and does not explicitly optimize jet-level observables; the resulting jet performance is therefore a consequence of improved particle reconstruction rather than a direct jet-level training objective.
Further improvements in jet energy resolution may be achieved through dedicated downstream jet \pt regression~\cite{CMS-DP-2024-064}.

In this work, the binary and PID classification thresholds in MLPF are not specifically optimized, and the balance between jet efficiency and purity is therefore expected to depend on the chosen threshold values.
We also note that both PF and MLPF have a stable jet resolution performance with respect to pileup within the range tested here (55--75 pileup interactions per event).
In the forward region, the jet resolution is generally comparable between PF and MLPF, while the efficiency is lower and the purity is higher for MLPF.
Moreover, tuning the model for the forward region is beyond the scope of this study, as the algorithm will need to be re-optimized for the HL-LHC era with the introduction of the high-granularity endcap calorimeter (HGCAL)~\cite{CMS:2017dhl} and the extended tracker coverage up to $\abs{\eta}<4$, which presents additional advantages for tracks that extrapolate into the HF region.

{\tolerance=1200
The performance of the \ptmiss reconstruction is evaluated by comparing the uncalibrated \ptmiss computed using PF or MLPF candidates, with the true, particle-level \ptmiss.
This comparison is shown in Fig.~\ref{fig:perf_met_nopu_cmssw} for both \ttbar events containing true \ptmiss from the primary event, and \qcd multijet events containing \ptmiss only through secondary processes such as \PQb decays or misreconstruction.
We find the \ptmiss distributions from PF and MLPF to be generally consistent, with MLPF reconstructing a somewhat harder \ptmiss spectrum in the tails.
\par
}

{\tolerance=1200
Since the model is trained using a particle-level target and does not explicitly optimize \ptmiss, the observed overall consistency, despite differences in the tails, indicates that particle-level optimization captures global physics properties of the event.
The \ptmiss resolution can be improved through dedicated downstream regression techniques, such as the \textsc{DeepMET} algorithm~\cite{CMS:DeepMET}, which takes PF candidates as input.
This work does not attempt to replace such techniques, but rather demonstrates that an ML-based reconstruction can provide alternative particle-level inputs to which \ptmiss regression methods can be applied.
\par
}

\section{Commissioning on data}
\label{sec:perf_data}

\begin{figure*}[htbp]
    \centering
    \includegraphics[width=0.32\linewidth]{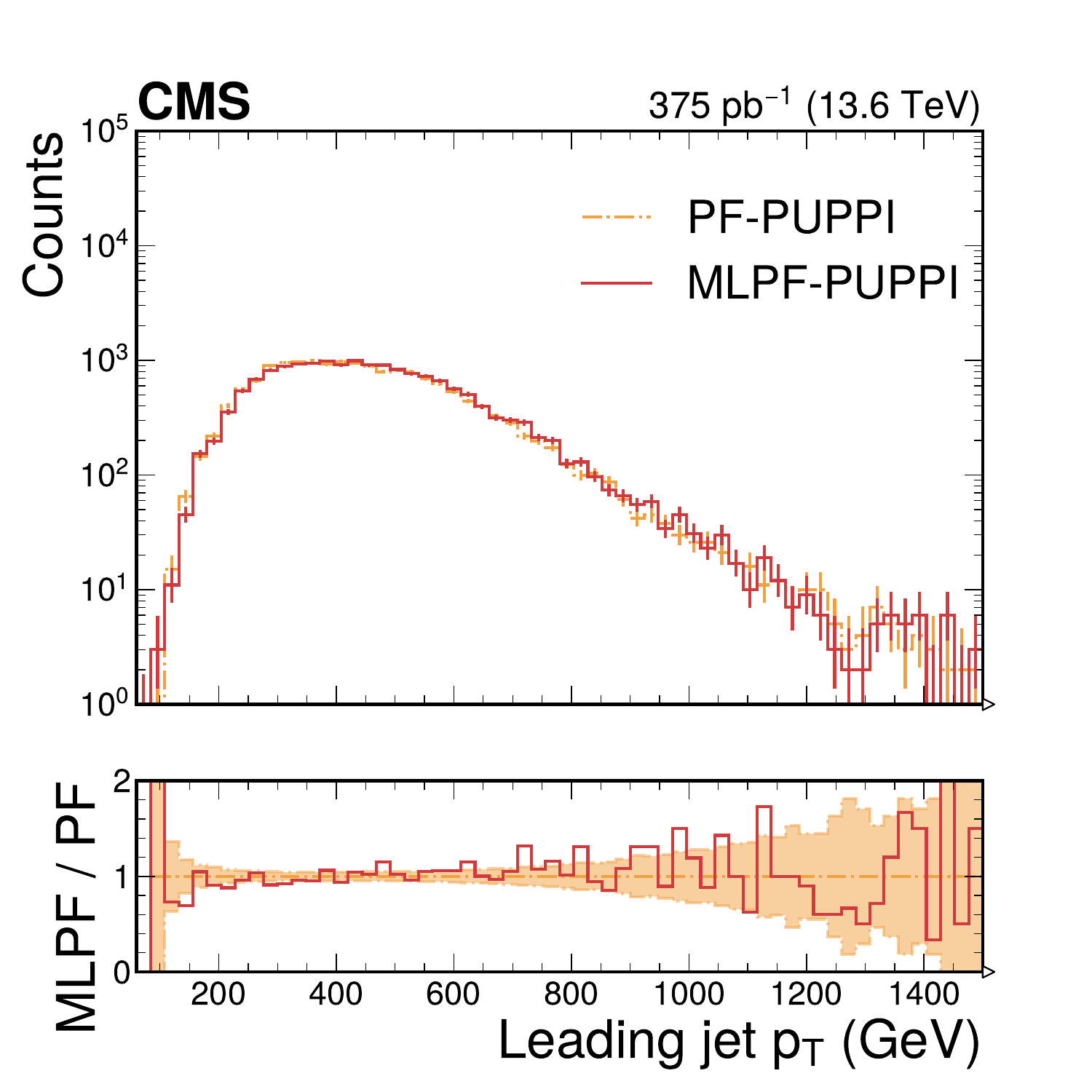}
    \includegraphics[width=0.32\linewidth]{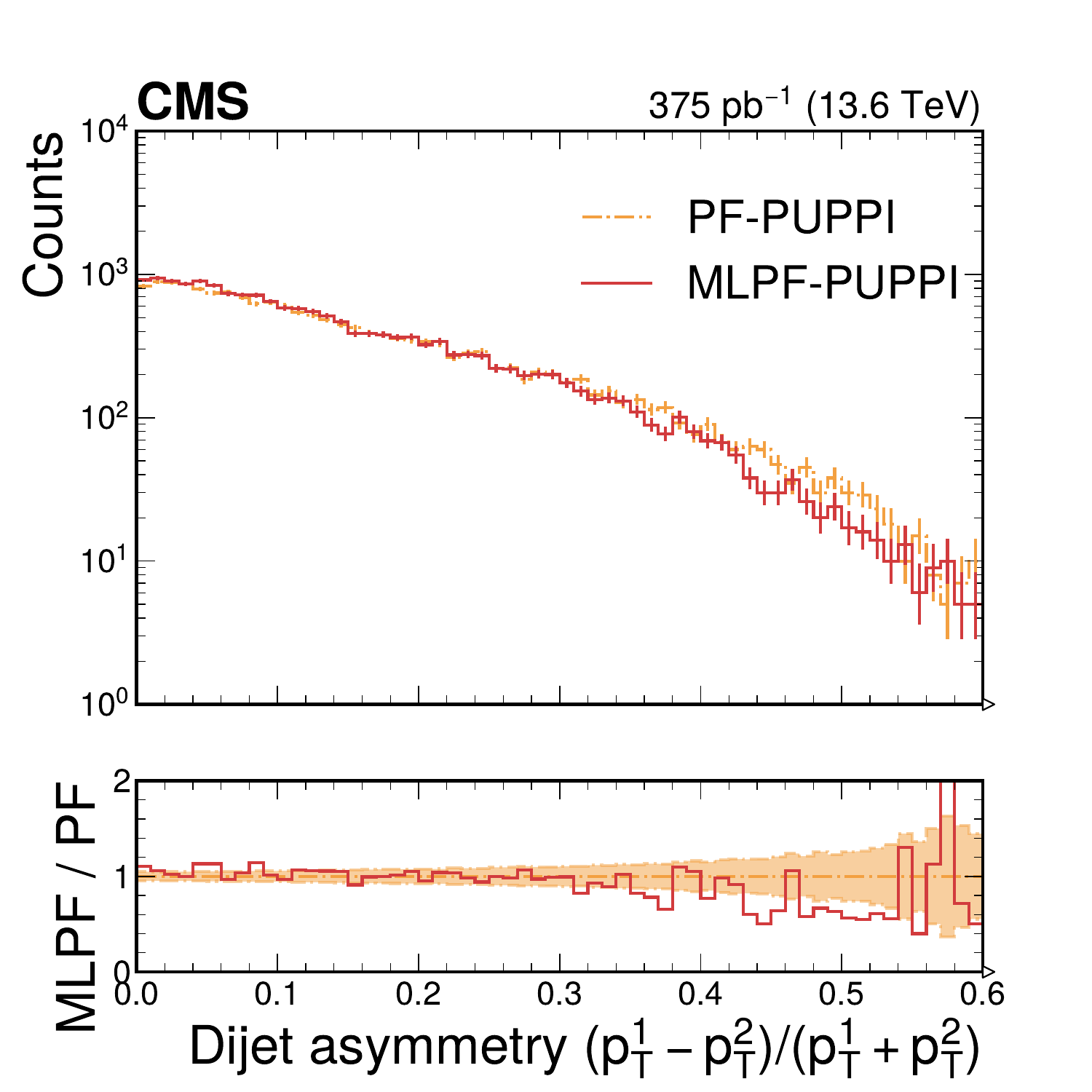}
    \includegraphics[width=0.32\linewidth]{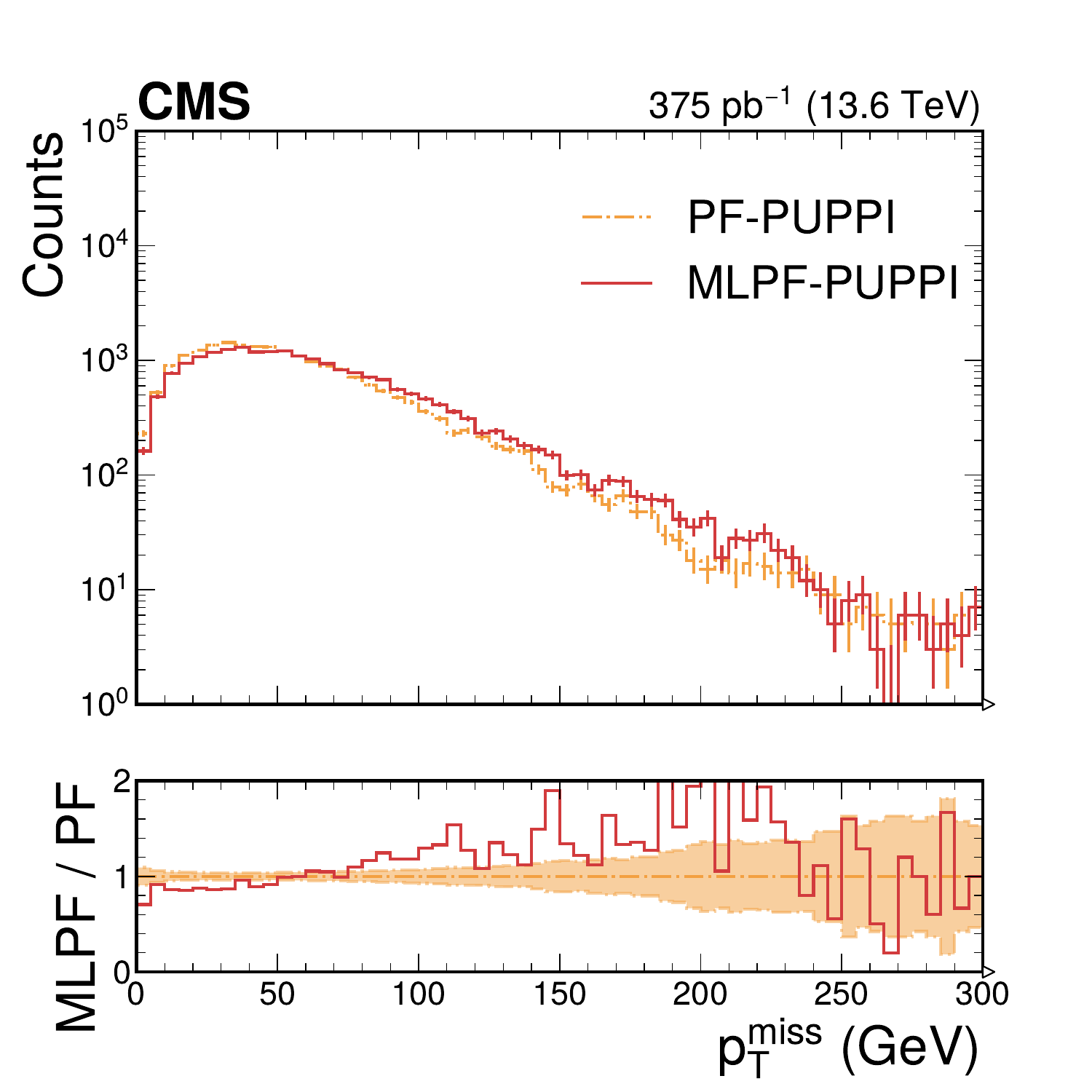}\\
    \caption{
        Distributions of the leading jet \ptCorr (left), the dijet asymmetry (center), and \ptmiss (right) in a small sample of dijet data collected in Run~3.
        The distributions are largely compatible.
        The vertical bars indicate the statistical uncertainties on the respective distributions, and the shaded bands in the ratio panel indicate the statistical uncertainties on PF.
    }
    \label{fig:perfdata}
\end{figure*}

The model was commissioned on a small sample of data collected in 2024 during Run~3, as described in Section~\ref{sec:datasets}.
We evaluate both PF (running on CPU) and MLPF (running on GPU) within the CMS offline software framework on the same dedicated machine.
The results are shown in Fig.~\ref{fig:perfdata}.
A dijet selection is applied to fully reconstructed events, requiring at least two central jets ($\abs{\eta}<2.5$) with $\pt > 60\GeV$ and an azimuthal separation of $\Delta \phi > 2.7$.
This selection ensures that we surpass the trigger efficiency turn-on region for the subleading jet \pt, which is required to have $\pt>45\GeV$ at the trigger level.
The jets are also required to pass the jet identification criteria intended to reject misidentified or poorly reconstructed jets based on the jet particle multiplicity, and neutral and charged-particle energy fractions~\cite{CMS:2017wyc}.
We find that MLPF and PF reconstruct comparable jet distributions in data.
The reconstructed dijet asymmetry distributions from PF and MLPF are found to be consistent, indicating that MLPF successfully reconstructs CMS data and produces realistic physics outputs.
Because the dijet asymmetry, in particular, is sensitive to the relative jet resolutions, this agreement further demonstrates that PF and MLPF achieve comparable jet-level performance.
The MLPF-reconstructed \ptmiss spectrum is somewhat harder, consistent with the trend observed in simulation (Fig.~\ref{fig:perf_met_nopu_cmssw}).
To deploy the MLPF algorithm in production, further validation and comparison of the re-reconstructed data with simulation is needed.

\section{Integration in the CMS reconstruction}
\label{sec:cmssw}

{\tolerance=1200
The MLPF model is exported to a static Open Neural Network Exchange (\ONNX)~\cite{onnx} file, freezing the weights and computational graph of the model.
This allows the model to be loaded via the \ONNXRUNTIME interface available in the CMS offline reconstruction framework.
To ensure efficient inference in events with thousands of PF elements, the \FlashAttention operation is exported as a fused kernel within the \ONNX graph, providing substantial memory and runtime improvements on Nvidia Ampere generation GPUs and newer.
Using the transformer-based model on full events generally requires an efficient \FlashAttention kernel implementation, but the slower and more memory-intensive fallback computation using naive matrix multiplication is possible when an efficient kernel implementation is not available.
At the time of writing, training with \FlashAttention is supported in \PYTORCH for both Nvidia (\eg, Ampere/\texttt{sm80} starting from \PYTORCH2.2) and AMD GPUs (\eg, MI200/\texttt{gfx90a} starting from \PYTORCH2.3).
Support for \FlashAttention during inference in \ONNXRUNTIME is currently only available for Nvidia GPUs (\eg, Ampere/\texttt{sm80}).
This is not a fundamental hardware limitation, but rather one of software support.
Implementations of \FlashAttention exist for other accelerators, \eg, Intel GPUs, but are currently only available in vendor-specific versions of \PYTORCH or \ONNXRUNTIME rather than in the main versions.
We have tested the inference on both Intel Xeon CPUs and Nvidia Ada Lovelace GPUs; broader hardware support (\eg, AMD GPUs, and ARM processors) is possible by improving the kernel implementations in \ONNXRUNTIME.
\par
}

The MLPF model is enabled in the CMS offline software via a configuration flag that replaces standard PF reconstruction with MLPF reconstruction, leaving other modules untouched.
The MLPF module takes tracks and clusters as inputs, performs inference through \ONNX, and converts the model outputs to a list of particle candidates per event.
Notably, the MLPF approach does not replace upstream tracking or cluster-building algorithms, which can efficiently be handled by existing GPU-based modules.
Subsequent reconstruction modules in the CMS offline software, such as those for jet clustering, charged hadron subtraction, and electron and tau reconstruction, can operate without modification using the PF candidates output by MLPF.
This integration enables the production of standard CMS analysis data tiers, allowing the model to be tested under realistic conditions, including on data.

\begin{figure}[bpht]
    \centering
    \ifthenelse{\boolean{cms@external}}
    {
        \includegraphics[width=\columnwidth]{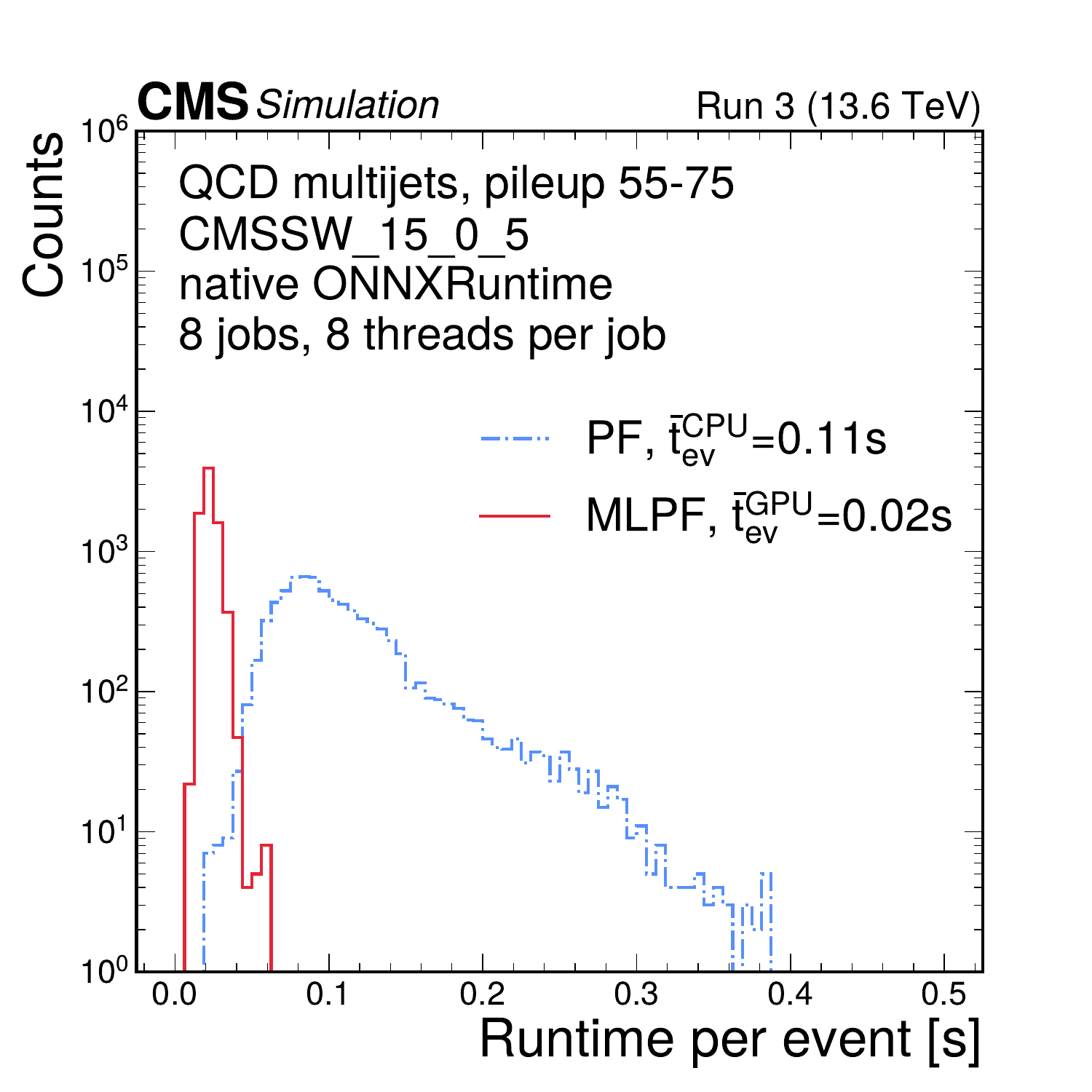}
        \includegraphics[width=\columnwidth]{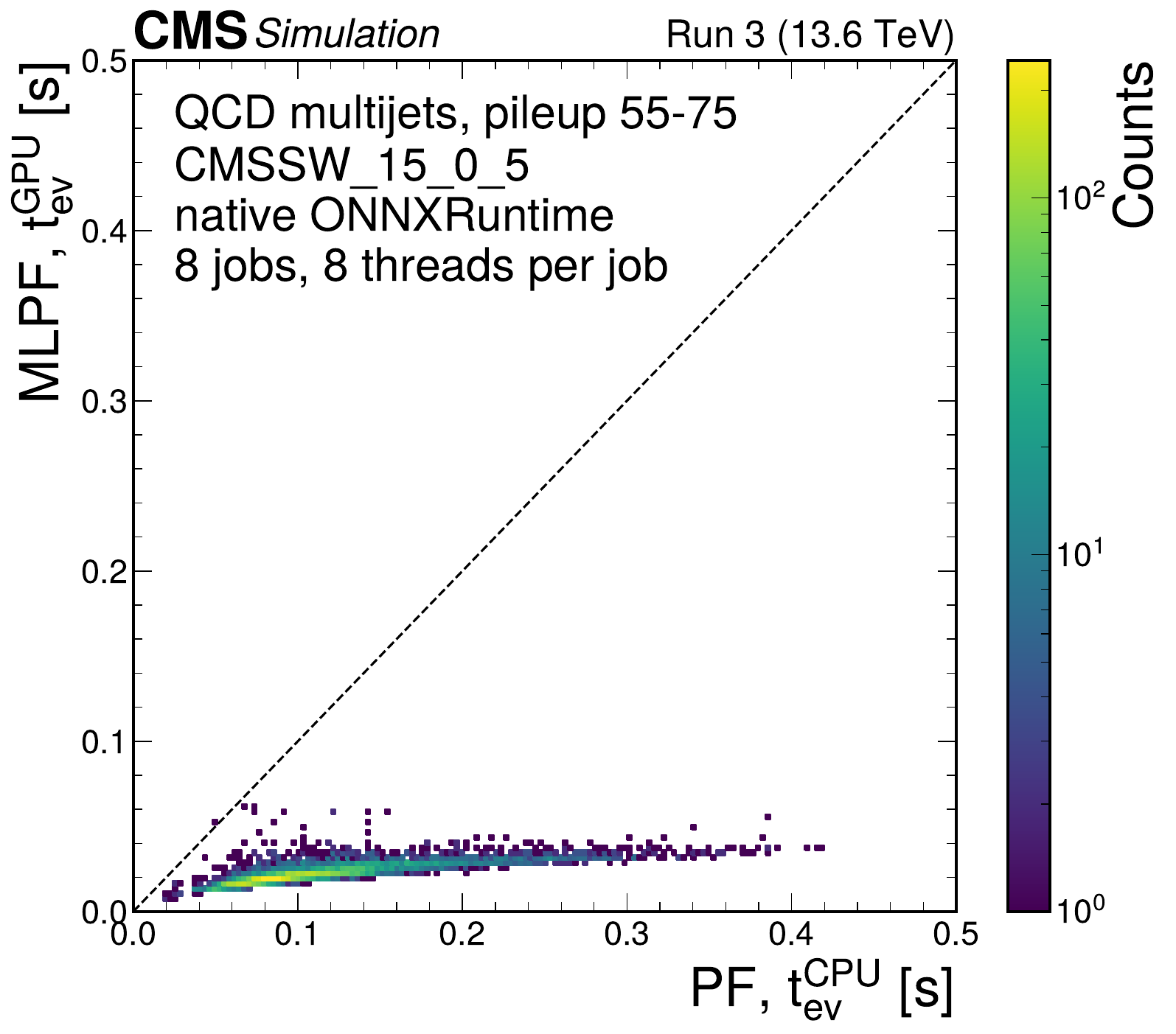}
    }
    {
        \includegraphics[width=0.47\linewidth]{Figure_010-a.pdf}
        \includegraphics[width=0.50\linewidth]{Figure_010-b.pdf}
    }
    \caption{
        The runtime of the baseline particle-flow reconstruction compared to the MLPF inference, directly within the CMS offline software using the native \ONNXRUNTIME on GPU.
    }
    \label{fig:onnx_runtime}
\end{figure}

Runtime performance was evaluated using eight parallel reconstruction jobs on a dual-socket Intel\textsuperscript{\tiny\textregistered} Xeon\textsuperscript{\tiny\textregistered} Platinum 8360Y CPU @ 2.40\unit{GHz} machine (144 hardware threads) sharing a single 24\unit{GB} Nvidia L4 GPU.
Each job ran eight threads, with each thread executing a separate inference stream.
Standard PF runs on CPUs, while MLPF performs inference on the GPU, executing synchronously on one event at a time.
In this configuration, one L4 GPU supported 64 parallel inference streams.
Figure~\ref{fig:onnx_runtime} shows the runtime performance of standard PF compared to the MLPF inference, directly within the CMS offline software using the native \ONNXRUNTIME on GPU, on Run-3 simulation samples with 55--75 simulated pileup interaction per event.
The runtime is measured for the complete PF reconstruction step by determining the time spent in the PF or MLPF module, including event setup, transfer to GPU, inference, transfer back to CPU, postprocessing the outputs to particles and storage in the event container.
The MLPF algorithm achieves an average runtime of 0.02 seconds per event, compared to 0.11 seconds per event for standard PF.
While the fraction of CPU time spent running PF with respect to the full offline reconstruction is relatively small under Run-3 conditions, with increasing event and detector complexity towards the HL-LHC, we expect the computational complexity of full event reconstruction to increase as well.
We note that the runtime per event is a simplified metric, and ultimately, one should benchmark full-node event throughput to assess if a given algorithm improves data processing efficiency.
This work shows that we can offload the PF linking and reconstruction step from CPU to GPU, thereby enabling computational accelerator usage for PF reconstruction with a meaningful expected throughput.

By correlating the event identifiers, we observe that the MLPF inference time remains narrow and stable across events, including those that are slow for PF, demonstrating improved runtime scaling with increasing event complexity, thanks in part to the \FlashAttention kernel.
Overall, GPU memory usage remains below 10\unit{GB} for 64 parallel inference streams (roughly 150\unit{MB} of GPU RAM per one inference stream).
Since the model is fairly small, we observe an average GPU compute utilization of 5--10\%.
Linearly scaling the memory usage, we extrapolate that one L4 GPU could support approximately 40 parallel inference threads at 200 pileup interactions without additional memory optimizations.
For reference, the inference runtime of the model on Intel Xeon Silver 4314 CPU @ 2.40\unit{GHz} using one thread is about a few seconds, demonstrating the necessity of using accelerators with optimized kernels for efficient ML inference in these circumstances.
Because of the iterative nature of the baseline PF block and particle reconstruction algorithms, a performant and fully-featured GPU implementation does not exist for PF, therefore MLPF unlocks the use of GPUs for this complex full-event reconstruction algorithm.

    {\tolerance=1200
        The current ML setup uses synchronous inference without event batching and full floating point precision, except for the \FlashAttention component, which runs in half precision.
        This leaves room for optimization through quantization, pruning, and asynchronous or batched execution~\cite{CMS:2024twn}.
        In this work, we evaluate the inference performance using an Nvidia L4 GPU, a successor to the T4 designed for power-efficient ML inference in data-center environments.
        We have also successfully used \FlashAttention for training on Nvidia A100, H100 GPUs as well as AMD MI250X GPUs, demonstrating the multi-vendor capability of the model.
        The MLPF model produces fully reproducible outputs: evaluating the same event multiple times on the same GPU yields identical results, and CPU and GPU inference agree within numerical precision.
        No stochastic components are used during inference, and no differences affecting physics performance are observed.
        \par}

\section{Summary}
\label{sec:sum}

{\tolerance=1200
A machine-learning (ML)-based algorithm for particle-flow (PF) reconstruction in the CMS experiment has been presented.
The algorithm correlates tracks and calorimeter clusters using successive transformer layers, and outputs a list of reconstructed particles with four-momenta and particle identification values.
The algorithm is functionally equivalent to the standard PF algorithm currently used in CMS.
The model is trained on CMS Run-3 (2023--2024) full simulation data sets that include realistic pileup conditions, and its performance is evaluated in simulated top quark-antiquark (\ttbar) and quantum chromodynamics multijet testing samples across different experimental conditions, as well as on data collected using a dijet trigger.
To ensure computational efficiency, the model avoids quadratic scaling in runtime and memory by efficiently employing modern graphics processing units (GPUs).
The ML-based PF (MLPF) algorithm achieves physics performance comparable to the standard PF algorithm, while enabling GPU acceleration for full event reconstruction within the CMS offline software framework.
Specifically, we observe improved neutral hadron efficiency with MLPF while maintaining the same misidentification rate as standard PF.
For jets with \pt between 30--100\GeV in \ttbar events, we observe a 10--20\% improvement in jet energy resolution compared to standard PF.
The algorithm delivers a per-event runtime of approximately 20\unit{ms} on an Nvidia L4 inference GPU, with 64 inference streams running in parallel per GPU.
We also observe improved runtime scaling with event size compared to the standard PF algorithm, highlighting the computational robustness of the approach.
\par
}

While the MLPF algorithm has been successfully benchmarked under Run-3 conditions, its primary motivation lies in deployment at the high-luminosity upgrade of the LHC (HL-LHC), where the increased event complexity and pileup conditions demand both high physics performance and computational scalability.
To deploy MLPF in production at the HL-LHC, several developments are foreseen.
First, the model must be retrained for the upgraded detector and higher pileup conditions.
Second, ensuring the portability of the inference across different GPU generations and hardware vendors is essential.
Third, the model inference can be further optimized through techniques such as sparse or block attention, quantization, pruning, and asynchronous or batched execution.
Finally, the physics performance and robustness can be improved through additional training and extensive cross-checks on a wide variety of simulated and data samples.

\begin{acknowledgments}
    We congratulate our colleagues in the CERN accelerator departments for the excellent performance of the LHC and thank the technical and administrative staffs at CERN and at other CMS institutes for their contributions to the success of the CMS effort. In addition, we gratefully acknowledge the computing centers and personnel of the Worldwide LHC Computing Grid and other centers for delivering so effectively the computing infrastructure essential to our analyses. Finally, we acknowledge the enduring support for the construction and operation of the LHC, the CMS detector, and the supporting computing infrastructure provided by the following funding agencies: SC (Armenia), BMBWF and FWF (Austria); FNRS and FWO (Belgium); CNPq, CAPES, FAPERJ, FAPERGS, and FAPESP (Brazil); MES and BNSF (Bulgaria); CERN; CAS, MoST, and NSFC (China); MINCIENCIAS (Colombia); MSES and CSF (Croatia); RIF (Cyprus); SENESCYT (Ecuador); ERC PRG, TARISTU24-TK10 and MoER TK202 (Estonia); Academy of Finland, MEC, and HIP (Finland); CEA and CNRS/IN2P3 (France); SRNSF (Georgia); BMFTR, DFG, and HGF (Germany); GSRI (Greece); NKFIH (Hungary); DAE and DST (India); IPM (Iran); SFI (Ireland); INFN (Italy); MSIT and NRF (Republic of Korea); MES (Latvia); LMTLT (Lithuania); MOE and UM (Malaysia); BUAP, CINVESTAV, CONACYT, LNS, SEP, and UASLP-FAI (Mexico); MOS (Montenegro); MBIE (New Zealand); PAEC (Pakistan); MES, NSC, and NAWA (Poland); FCT (Portugal);  MESTD (Serbia); MICIU/AEI and PCTI (Spain); MOSTR (Sri Lanka); Swiss Funding Agencies (Switzerland); MST (Taipei); MHESI (Thailand); TUBITAK and TENMAK (T\"{u}rkiye); NASU (Ukraine); STFC (United Kingdom); DOE and NSF (USA).

    \hyphenation{Rachada-pisek} Individuals have received support from the Marie-Curie program and the European Research Council and Horizon 2020 Grant, contract Nos.\ 675440, 724704, 752730, 758316, 765710, 824093, 101115353, 101002207, 101001205, and COST Action CA16108 (European Union); the Leventis Foundation; the Alfred P.\ Sloan Foundation; the Alexander von Humboldt Foundation; the Science Committee, project no. 22rl-037 (Armenia); the Fonds pour la Formation \`a la Recherche dans l'Industrie et dans l'Agriculture (FRIA) and Fonds voor Wetenschappelijk Onderzoek contract No. 1228724N (Belgium); the Beijing Municipal Science \& Technology Commission, No. Z191100007219010, the Fundamental Research Funds for the Central Universities, the Ministry of Science and Technology of China under Grant No. 2023YFA1605804, the Natural Science Foundation of China under Grant No. 12535004, and USTC Research Funds of the Double First-Class Initiative No.\ YD2030002017 (China); the Ministry of Education, Youth and Sports (MEYS) of the Czech Republic; the Shota Rustaveli National Science Foundation, grant FR-22-985 (Georgia); the Deutsche Forschungsgemeinschaft (DFG), among others, under Germany's Excellence Strategy -- EXC 2121 ``Quantum Universe" -- 390833306, and under project number 400140256 - GRK2497; the Hellenic Foundation for Research and Innovation (HFRI), Project Number 2288 (Greece); the Hungarian Academy of Sciences, the New National Excellence Program - \'UNKP, the NKFIH research grants K 131991, K 133046, K 138136, K 143460, K 143477, K 146913, K 146914, K 147048, 2020-2.2.1-ED-2021-00181, TKP2021-NKTA-64, and 2025-1.1.5-NEMZ\_KI-2025-00004 (Hungary); the Council of Science and Industrial Research, India; ICSC -- National Research Center for High Performance Computing, Big Data and Quantum Computing, FAIR -- Future Artificial Intelligence Research, and CUP I53D23001070006 (Mission 4 Component 1), funded by the NextGenerationEU program (Italy); the Latvian Council of Science; the Ministry of Education and Science, project no. 2022/WK/14, and the National Science Center, contracts Opus 2021/41/B/ST2/01369, 2021/43/B/ST2/01552, 2023/49/B/ST2/03273, and the NAWA contract BPN/PPO/2021/1/00011 (Poland); the Funda\c{c}\~ao para a Ci\^encia e a Tecnologia, grant CEECIND/01334/2018 (Portugal); the National Priorities Research Program by Qatar National Research Fund; MICIU/AEI/10.13039/501100011033, ERDF/EU, ``European Union NextGenerationEU/PRTR", and Programa Severo Ochoa del Principado de Asturias (Spain); the Chulalongkorn Academic into Its 2nd Century Project Advancement Project, the National Science, Research and Innovation Fund program IND\_FF\_68\_369\_2300\_097, and the Program Management Unit for Human Resources \& Institutional Development, Research and Innovation, grant B39G680009 (Thailand); the Kavli Foundation; the Nvidia Corporation; the SuperMicro Corporation; the Welch Foundation, contract C-1845; and the Weston Havens Foundation (USA).
\end{acknowledgments}

\bibliography{auto_generated}

\cleardoublepage \appendix\section{The CMS Collaboration \label{app:collab}}\begin{sloppypar}\hyphenpenalty=5000\widowpenalty=500\clubpenalty=5000\input{PFT-25-001-public-authorlist.tex}\end{sloppypar}
\end{document}

%% file: PFT-25-001-public-authorlist.tex
\cmsinstitute{Yerevan Physics Institute, Yerevan, Armenia}
{\tolerance=6000
A.~Hayrapetyan, V.~Makarenko\cmsorcid{0000-0002-8406-8605}, A.~Tumasyan\cmsAuthorMark{1}\cmsorcid{0009-0000-0684-6742}
\par}
\cmsinstitute{Institut f\"{u}r Hochenergiephysik, Vienna, Austria}
{\tolerance=6000
W.~Adam\cmsorcid{0000-0001-9099-4341}, L.~Benato\cmsorcid{0000-0001-5135-7489}, T.~Bergauer\cmsorcid{0000-0002-5786-0293}, M.~Dragicevic\cmsorcid{0000-0003-1967-6783}, P.S.~Hussain\cmsorcid{0000-0002-4825-5278}, M.~Jeitler\cmsAuthorMark{2}\cmsorcid{0000-0002-5141-9560}, N.~Krammer\cmsorcid{0000-0002-0548-0985}, A.~Li\cmsorcid{0000-0002-4547-116X}, D.~Liko\cmsorcid{0000-0002-3380-473X}, M.~Matthewman, J.~Schieck\cmsAuthorMark{2}\cmsorcid{0000-0002-1058-8093}, R.~Sch\"{o}fbeck\cmsAuthorMark{2}\cmsorcid{0000-0002-2332-8784}, M.~Shooshtari\cmsorcid{0009-0004-8882-4887}, M.~Sonawane\cmsorcid{0000-0003-0510-7010}, W.~Waltenberger\cmsorcid{0000-0002-6215-7228}, C.-E.~Wulz\cmsAuthorMark{2}\cmsorcid{0000-0001-9226-5812}
% ...
\par}
\vskip\cmsinstskip
\dag:~Deceased\\
$^{1}$Also at Yerevan State University, Yerevan, Armenia\\
$^{2}$Also at TU Wien, Vienna, Austria\\
% ...
